\documentclass{aa}
\usepackage{graphicx}
\usepackage{txfonts}
\usepackage{natbib}
\usepackage{color}
\bibpunct{(}{)}{;}{a}{}{,}    %to follow the A&A style 
\newcommand{\beq}{\begin{equation}}
\newcommand{\eeq}{\end{equation}}
\newcommand{\bea}{\begin{eqnarray}}
\newcommand{\eea}{\end{eqnarray}}
\newcommand{\url}[1]{{\tt #1}}

\def\gapp{\lower 3pt\hbox{${\buildrel > \over \sim}$}\ }
\def\lapp{\lower 3pt\hbox{${\buildrel < \over \sim}$}\ }

\usepackage{color}
\usepackage{ulem}

\newlength{\linwx}
\begin{document}
\title{Stellar irradiated discs and implications on migration of embedded planets III: viscosity transitions}
%  \subtitle{empty}
%%
\author{
Bertram Bitsch \inst{1,2},
Alessandro Morbidelli \inst{2},
Elena Lega \inst{2},
Katherine Kretke \inst{3},
Aur\'{e}lien Crida \inst{2}
}
\offprints{B. Bitsch,\\ \email{bert@astro.lu.se}}
\institute{
Lund Observatory, Department of Astronomy and Theoretical Physics, Lund University, Box 43, 22100 Lund, Sweden
\and
Laboratoire Lagrange, UMR7293, Universit\'{e} Nice Sophia-antipolis / CNRS / Observatoire de la C\^{o}te d'Azur, 06300 Nice, France
%University Nice-Sophia Antipolis, CNRS, Observatoire de la C\^{o}te d'Azur,
%Laboratoire LAGRANGE, CS 34229, 06304 NICE cedex 4, FRANCE
\and
Southwest Research Institute, Department of Space Studies, 1050 Walnut Street, Suite 300, Boulder, CO 80302, USA
}
\abstract
%%  Context
{ The migration strength and direction of embedded low-mass planets depends on the disc structure. In discs with an efficient radiative transport, the migration can be directed outwards for planets with more than $3-5$ Earth masses. This is due to the entropy driven corotation torque, a process that extends the lifetimes of growing planetary embryos. However, smaller mass planets are still migrating inwards and might be lost to the central star. 
 } 
%%  Aims
{ We investigate the influence on the disc structure caused by a jump in the $\alpha$ parameter of the viscosity to model a dead-zone structure in the disc. We focus on $\dot{M}$ discs, which have a constant net mass flux. Using the resulting disc structure, we investigate the consequences for the formation of planetesimals and determine the regions of outward migration for proto-planets.
} 
%% Methods
{ We performed numerical hydrosimulations of $\dot{M}$ discs in the r-z-plane. We used the explicit/implicit hydrodynamical code FARGOCA that includes a full tensor viscosity and stellar irradiation as well as a two-temperature solver that includes radiation transport in the flux-limited diffusion approximation. The migration of embedded planets was studied by using torque formulae.
} 
%% Results
{
Viscosity transitions inside the disc create transitions in density that stop inward migration for small planets through the so-called "planet trap" mechanism. This mechanism also works for planets down to $M_P > 0.5M_{Earth}$, while in radiative discs with no viscosity transition the lowest mass with which inward migration can be avoided is $3-5$ Earth masses. Additionally, the viscosity transitions change the pressure gradient in the disc, which facilitates planetesimal formation via the streaming instability. However, a very steep transition in viscosity is needed to achieve in a pressure bump in the disc.
}
%% Conclusions
{
The transition in viscosity facilitates planetesimal formation and can stop the migration of small-mass planets ($M_P>0.5M_{Earth}$), but still does not halt inward migration of smaller planets and planetesimals that are affected by gas drag. A very steep, probably unrealistic viscosity gradient is needed to trap planets of smaller masses and halt gas-drag-driven planetesimal migration at a pressure bump.
}
\maketitle
\markboth
{Bitsch et al.: Stellar irradiated discs and implications on migration of embedded planets}
{Bitsch et al.: Stellar irradiated discs and implications on migration of embedded planets}

\section{Introduction}
\label{sec:introduction}

Planet formation occurs in accretion discs around young stars. The evolution of accretion discs is still not understood completely, but it takes place on a Myr time-scale during which the accretion rate of gas onto the host star drops \citep{1998ApJ...495..385H}. The formation of gas giants has to take place during this time. In the most commonly favoured scenario \citep{1996Icar..124...62P} a massive solid core of several Earth masses has to form first to accrete gas. However, the formation of cores via the accretion of embryos and planetesimals is not very efficient and therefore can not explain the growth of proto-planetary cores \citep{2010AJ....139.1297L}. Recently, the idea of pebble accretion onto planetary embryos offered a possible solution to this problem, because it occurs on a very short time-scale \citep{2012A&A...544A..32L, 2012A&A...546A..18M}. 

Regardless of how the cores may form, they will migrate through the disc because of interactions with the gas \citep{1997Icar..126..261W}. As long as discs had been assumed to be locally isothermal, cores were expected to migrate very fast towards the central star, which was a problem for understanding the formation of giant planets. It has been recently shown that the migration of the planetary cores in the disc depends on the thermodynamics in the disc \citep{2006A&A...459L..17P, 2008A&A...487L...9K, 2008ApJ...672.1054B, 2009A&A...506..971K}. For a review on planet migration see \citet{2013arXiv1312.4293B}. The thermodynamics inside the discs are determined by viscous heating, radiative cooling, and stellar illumination (\citet{2013A&A...549A.124B}, hereafter Paper I). Stellar irradiation maintains a flared disc structure in the outer parts of the disc (\citet{1997ApJ...490..368C}; Paper I). A consequence of the disc structure on the migration of planets is described in Paper I, where we find that outward migration is possible in the regions of the disc where $H/r$ decreases, which effectively means that the regions of outward migration are smaller in a flared disc than a shadowed disc, where $H/r$ decreases in the outer parts of the disc.

\citet{2014arXiv1401.1334B} hereafter Paper II studied the evolution of discs where the mass flow towards the star ($\dot{M}$) is independent of radius. In these accretion discs, the mass flux $\dot{M}$ is defined as 
\begin{equation}
 \dot{M} =  3 \pi \nu \Sigma_G = \alpha H^2 \Omega_K \Sigma_G \ ,
\end{equation}
where $\nu$ is the viscosity, following the $\alpha$-viscosity approach \citep{1973A&A....24..337S}. $H$ is the height of the disc, and $\Omega_K$ is the Keplerian rotation frequency. $\Sigma_G$ denotes the gas surface density. We have found that during the evolution of the disc (decreasing $\dot{M}$) the regions of outward migration shrink until for small $\dot{M}$ no outward migration is possible. Consequently, all planets migrate inwards, similarly to the isothermal disc scenario. This, again, may be problematic for understanding giant planet formation. Moreover, even when regions of outward migration exist, they typically concern only planets more massive than $\approx 5$ Earth masses. Smaller planets migrate inwards as they grow, and therefore it is debated whether they can achieve this threshold mass quickly enough to be captured in the region of outward migration. Additionally, the inward migration of small-mass planets might be even faster than estimated in \citet{2011MNRAS.410..293P}, see \citet{Lega2013}. All this indicates that we still miss a mechanism for preventing inward migration of low-mass protoplanets.

The inward migration of small-mass planets can be stopped by a bump in vortensity \citep{2006ApJ...642..478M}, which is defined as $(\nabla \times v) / \Sigma $, where $v$ is the velocity of the gas. The bump in vortensity (positive radial gradient) results in a positive contribution of the barotropic part of the corotation torque that acts on the planet which can overcompensate for the negative Lindblad torque and halt its inward migration.

Additionally, the formation of planetesimals and planetary embryos can be aided by a positive radial pressure gradient in a protoplanetary disc \citep{2007ApJ...662..627J}. This stops the inward migration of aggregates that undergo gas drag \citep{2008A&A...487L...1B} and significantly facilitates planetesimal formation \citep{2010ApJ...722.1437B, 2010ApJ...722L.220B}. This indicates that a bump in vortensity and pressure would be very useful for the formation of planetesimals and planetary embryos by the core accretion scenario, which raises the question of how such a vortensity bump and a pressure bump can appear in the disc.

The radial changes of the vortensity and pressure fields can be triggered by changes in viscosity, which can be caused by changes in the gas ionisation fraction or condensation at the ice line (see sect~\ref{subsec:deadzone}). In Paper II we only explored discs that have a unique value of $\alpha$ in time and space. In this paper, we investigate the effect of viscosity transitions on the structure of the disc, with special attention devoted to the radial gradients of vortensity and pressure. Our aim is to determine the conditions for the following appealing scenario for giant planet formation:

At a pressure bump in the disc, particles can accumulate and form planetesimals and planetary embryos. These embryos would then migrate inwards, therefore we assume that a bump in vortensity exists at the same location in the disc, which can trap the planetary embryos \citep{2006ApJ...642..478M, 2008A&A...478..929M}. After they are formed, these embryos could then grow via pebble accretion \citep{2012A&A...544A..32L, 2012A&A...546A..18M} until they reach a mass of $\approx 10-20M_{Earth}$ and gas accretion onto the core can start \citep{1996Icar..124...62P}. The inward migration of cores of a few Earth masses can also be stopped by the entropy related corotation torque \citep{2006A&A...459L..17P, 2008A&A...487L...9K, 2008ApJ...672.1054B, 2009A&A...506..971K}, so that at this stage of evolution a bump in vortensity is not necessarily needed any more. As the planet starts to grow beyond $\approx 50M_{Earth}$, it opens up a gap in the disc \citep{2006Icar..181..587C} and is then released into type-II-migration \citep{1986ApJ...307..395L} and slowly moves towards the star as the disc is accreted. The final stopping point of the type-II migrating gas giant planet is then set by the inner edge of the disc \citep{2007MNRAS.377.1324C} or the photo-evaporation radius of the disc \citep{2012MNRAS.422L..82A}, which explains the pile-up of large gas giants in the inner systems.

This paper is structured as follows: First, we give an overview of the methods used, especially of the viscosity (section~\ref{sec:methods}). Then we compare in great detail a $\dot{M}=1\times 10^{-7}M_\odot/yr$ disc with and without transition in viscosity in section~\ref{sec:visctrans}. In section~\ref{sec:discevolve} we investigate the changes of the disc structure as the disc evolves to smaller $\dot{M}$. Different variations of viscosity transitions are then discussed in section~\ref{sec:difftrans}. In section~\ref{sec:presgrad} we discuss which viscosity transition is actually needed to realise our planet formation scenario, and we discuss how realistic it might be. Finally, we give a summary in section~\ref{sec:summary}.

\section{Methods}
\label{sec:methods}

\subsection{General setup}

The general simulation setup follows the descriptions of Papers I and II. The protoplanetary gas disc is treated as a three-dimensional (3D) non-self gravitating gas whose motion is described by the Navier-Stokes equations. Without any perturbers in the disc, the disc has an axisymmetric structure. We therefore can compute all quantities in 2D, in the $r$-$\theta$-plane, where $\theta$ is the colatitude of the spherical coordinates centred on the star, such that $\theta=90^\circ$ in the midplane. We used a $386 \times 66$ grid cells for all simulations. The dissipative effects are described via the standard viscous stress tensor approach \citep[e.g.][]{1984frh..book.....M}. We also included the irradiation from the central star, which was described in detail in Paper I. For that purpose we modified and substantially extended the existing multi-dimensional hydrodynamical code FARGOCA, as presented in \citet{Lega2013} and Paper II.

Viscous and stellar heating generates thermal energy in the disc, which diffuses radiatively through the disc and is emitted from its surfaces. To describe this process we used the flux-limited diffusion approximation \citep[FLD,][]{1981ApJ...248..321L}, an approximation that allows the transition from the optically thick mid-plane to the thin regions near the disc's surface; more details of this are given in Paper I. For comparison purposes with Paper II, we used a star with $M_\star = 1 M_\odot$ and $R_\star = 3R_\odot$ and $T_\star = 5600$K, which corresponds to $L\approx 9 L_\odot$. The central star mainly influences the parts of the disc that are dominated by stellar irradiation. For the inner disc, we do not assume that this plays a significant role.

We adopted the opacity law of \citet{1994ApJ...427..987B}, with $\kappa_\star=3.5{\rm cm}^2/{\rm g}$ as in Paper II. Throughout all the simulations presented here, we used a metallicity of $1\%$ in $\mu m$ size dust grains. We also apply the $\dot{M}$ boundary conditions described in the appendix of Paper II.

\subsection{Prescription of viscosity transitions}
\label{subsec:deadzone}

Here we explore the effects of putative viscosity transitions at the ice line on the formation of planetesimals via the streaming instability as well as their capability of halting the migration of small planets. More precisely, we are interested in the viscosity gradients (that change the gradients in surface density and pressure) necessary to support our proposed scenario of planet formation. We modelled the transition in viscosity in a simple toy model, instead of using full-scale {\bf M}agneto-{\bf R}otational-{\bf I}nstability (MRI) simulations \citep{1998RvMP...70....1B}. We made the general assumption that the $\alpha$-parameter of our viscosity \citep{1973A&A....24..337S} in the active layer ($\alpha_A$) is much larger than in the midplane regions of the disc, which corresponds to a dead zone like structure \citep{1998RvMP...70....1B}. 

\citet{2010A&A...520A..14P} have studied the influence of a dead zone on the accretion of gas onto giant cores in isothermal discs. They modelled the dead zone by an $\alpha$ transition in the vertical direction dependent on $H$ in the disc. They found that the gas accretion process is not sufficiently altered by the presence of a dead zone, but that inward migration is somewhat reduced for giant planets. Here we instead focus on the migration of small-mass planets, and moreover, we link the transitions in $\alpha$ to the temperature and column density in the disc.

We took two main boundaries for the $\alpha$ transition into account. (i) An inner radial boundary that is dependent on the temperature (thermal ionisation), and (ii) a boundary that depends on the vertically integrated volume density (due to the absorption of X-ray and/or cosmic rays). In the midplane regions of the disc, we then explored the influence of an additional transition of viscosity at the ice line, where the different amount of condensed dust/ice grains there can quench the MRI at different degrees \citep{2007ApJ...664L..55K}. The idea is not to give a priori quantitative estimates for these transition, but to understand the general behaviour of a disc that has viscosity transitions. Therefore the prescriptions also are slightly modified to test different parameters in section~\ref{sec:difftrans}.

The general assumption is that for $T>T_{crit}=1000$K the disc is fully ionised and therefore MRI active. This would correspond to the inner parts of the disc, and maybe to some super-heated parts in the atmosphere of the accretion disc. For $T<T_{crit}$ the MRI activity has to be triggered by different sources. At the upper and lower layers of the disc, ionization originates from cosmic rays, stellar X-rays, ultraviolet photons and radionuclides. All these ionization sources have different penetration depths into the disc and therefore ionize the disc down to different levels. 

Ultraviolet photons will only penetrate to a mass column of $0.1g/cm^2$ \citep{2011ApJ...727....2P}. X-rays can contribute down to $\Sigma=10g/cm^2$ \citep{2009ApJ...701..737B}, while cosmic rays can penetrate down to $\Sigma=100g/cm^2$ \citep{2013ApJ...765..114D}. The different penetration depths for cosmic rays and X-rays suggests that an additional transition of $\alpha$ exists in the vertically integrated volume density.

Additionally, we imposed a transition of $\alpha$ at the ice line because there might be ice grains below the water condensation temperature that can then quench the MRI more effectively than in the region where only silicate grains are available \citep{2007ApJ...664L..55K}. For the different $\alpha$ values we used we defined the following relation:
\begin{equation}
\label{eq:alpharatio}
 \alpha_A = 10 \alpha_U = 100 \alpha_D \ ,
\end{equation}
where we set $\alpha_A = 0.0264$. This indicates that $\alpha_U= 0.00264$ ($\alpha$ undead), which is still a significant, but low, alpha factor. In the $\alpha_D$ ($\alpha$ dead) region, the $\alpha$ factor is even smaller. 

A zero viscosity in the dead zone can lead to outbursts in the disc triggered by gravitationally instabilities in the disc \citep{2014MNRAS.437..682M}. Inside the dead zone, however, radionuclides can provide an additional source of ionization, and hydrodynamical instabilities might create turbulence, for instance, the vertical shear instability \citep{2013MNRAS.435.2610N} or the baroclinic instability \citep{2003ApJ...582..869K}, which will provide some viscosity. Therefore we assumed that $\alpha_D$ is not zero.

We set the following $\alpha$ regions in the disc:
\begin{eqnarray}
\label{eq:alpha}
 \Sigma_P (z) \leq 10g/cm^2 \quad {\rm OR} \quad 1000K \leq T \quad &\Rightarrow& \quad \alpha_A \nonumber \\
 10g/cm^2 < \Sigma_P (z) \leq 50g/cm^2 \quad &\Rightarrow& \quad \alpha_A \to \alpha_U \nonumber \\
 50g/cm^2 < \Sigma_P (z); 220K \leq T<800K \quad &\Rightarrow& \quad \alpha_U \nonumber \\
 50g/cm^2 < \Sigma_P (z); 160K \leq T<220K \quad &\Rightarrow& \quad \alpha_U \to \alpha_D \nonumber \\
 50g/cm^2 < \Sigma_P (z) \leq 90g/cm^2; T<160K \quad &\Rightarrow& \quad \alpha_U \to \alpha_D \nonumber \\
 90g/cm^2 < \Sigma_P (z); T<160K \quad &\Rightarrow& \quad \alpha_D \nonumber \\
 800K \leq T < 1000K \quad &\Rightarrow& \quad \alpha_A \to \alpha_U \ ,
\end{eqnarray}
where $\to$ symbolizes a transition region for $\alpha$. $\Sigma_P (z)$ is the vertically integrated volume density from the top of the disc down to a height $z$, where the desired quantity of $\Sigma_P$ is reached. This basically represents the penetration depth of cosmic and X-rays. We obtained the gas surface density by integrating both sides of the disc from top to midplane and from bottom to midplane. Therefore the gas surface density of the disc $\Sigma_G$ is $2\Sigma_P (0)$. Transitions for the different $\alpha$ regions need to be smoothed out, because this is closer to real MHD simulations than a step function \citep{2013ApJ...765..114D}.

\begin{figure}
 \centering
 \includegraphics[scale=0.78]{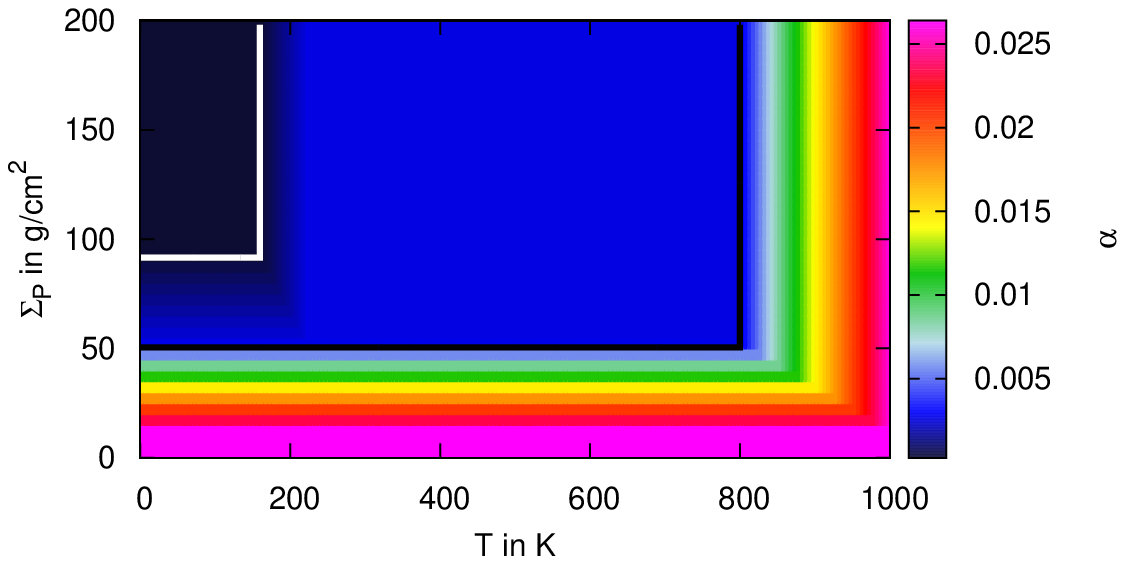}
 \includegraphics[scale=0.78]{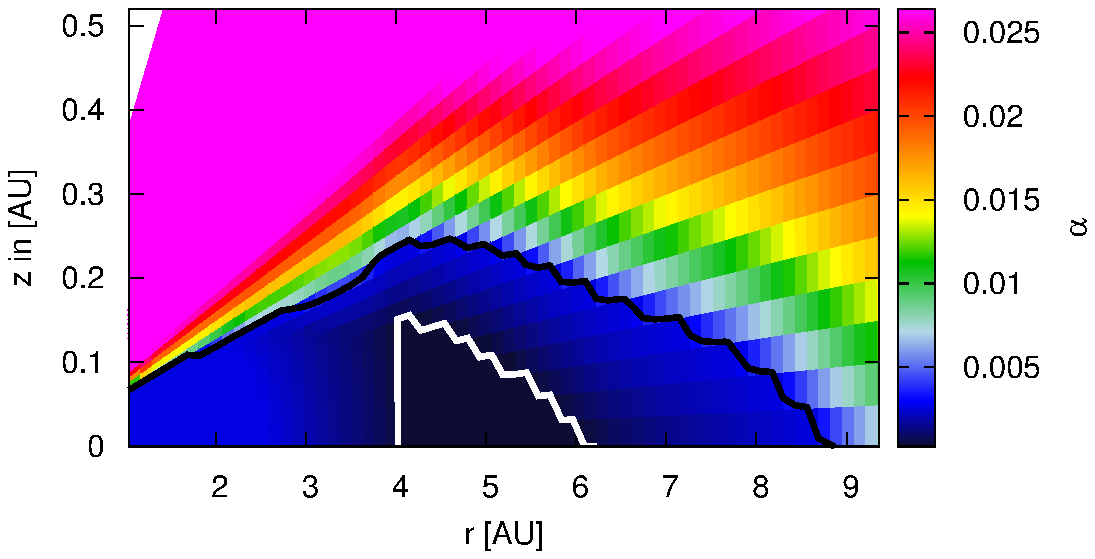}
 \caption{{\bf Top:} $\alpha$ as a function of $\Sigma_P$ and $T$ as specified by eq.~\ref{eq:alpha}. {\bf Bottom:} $\alpha$ parameter in an $\dot{M}=1\times 10^{-7}M_\odot/yr$ disc that is discussed in section~\ref{sec:visctrans}. Inside the black line $\alpha \leq \alpha_U$, inside the white line $\alpha = \alpha_D$.
   \label{fig:Alpha}
   }
\end{figure}

Note here that if either $T>1000$K or $\Sigma_P(z) <10$g/cm$^2$, the $\alpha$ value corresponds to $\alpha_A$, because the disc is either completely ionised by thermal ionisation or by X-rays and cosmic rays. The transition regions are smoothed in the following, linear way:
\begin{eqnarray}
 \alpha_T &= \alpha_A - (\alpha_A - \alpha_U) \frac{\Sigma_P-\Sigma_A}{40g/cm^2} \quad {\rm for} \quad \alpha_A \to \alpha_U \\
 \alpha_T &= \alpha_A - (\alpha_A - \alpha_U) \frac{1000K-T}{200K} \quad {\rm for} \quad \alpha_A \to \alpha_U \ ,
\end{eqnarray}
and likewise for the other transition regions. Parts of the disc that are in these transition regions either follow the gradient of $\Sigma_P$ or of $T$, depending on the steepness of these gradients (where the shallower gradient is favoured). Note here that the temperature range over which we imposed the $\alpha$-transition ($\alpha_U \to \alpha_D$) at the ice line is exactly the range over which the opacity changes in the opacity we used \citep{1994ApJ...427..987B}. We therefore mark in our plots the ice line at $190$K, which is the middle of that transition range. The viscosity is given by
\begin{equation}
 \nu = \alpha c_s^2 / \Omega_K \ ,
\end{equation}
where $\Omega_K$ is the Keplerian frequency and $c_s$ is the sound speed. Here we used the midplane sound speed $c_s (z=0)$, which provides a better control over the viscosity, which varies with height only because of changes in the $\alpha$ parameter. Additionally, the vertical changes in $c_s$ are expected to be much smaller than the changes in $\alpha$. The $\alpha$ map as a function of $\Sigma_P$ and $T$ is shown in Fig.~\ref{fig:Alpha} (top). We also display the $\alpha$ map of an $\dot{M}=1\times 10^{-7}M_\odot/yr$ disc (bottom panel) in equilibrium state (constant mass flux through each radial section of the disc) with $\alpha_A = 0.0264$. The viscosity used for $\alpha_A$ in this work is generally ten times higher than in Paper II. 

\section{Viscosity transitions}
\label{sec:visctrans}

In Paper II we focused on discs with constant $\dot{M}$. These discs also featured a constant $\alpha$ parameter in the whole disc. As stated above, the $\alpha$ parameter of the viscosity now follows some transitions, so that $\alpha$ changes in the disc with $r$ and $z$ because the temperature and the volume density are also functions of $r$ and $z$. The radial and vertical changes in $\alpha$ will lead to changes in volume density relative to the uniform $\alpha$ case, because $\dot{M}$ has to be independent of $r$ in the steady state. We compare here a disc with the described transition in $\alpha$ with a disc that has constant $\alpha$ throughout the disc equal to the value of $\alpha_A$.

\subsection{Disc structure}
\label{subsec:structure}

\begin{figure}
 \centering
 \includegraphics[scale=0.69]{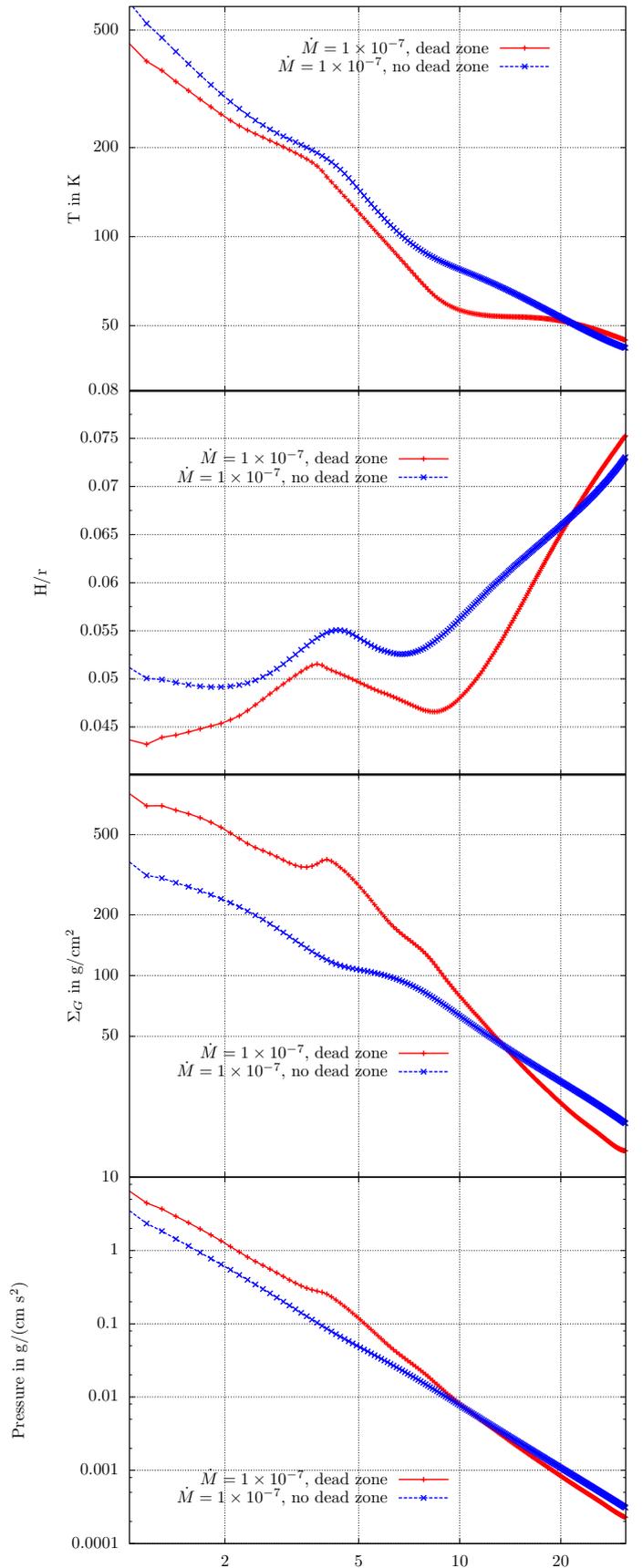}
 \caption{Midplane temperature (top), $H/r$ (second from top), integrated surface density $\Sigma_G$ (third from top) and midplane pressure (bottom) for the $\dot{M}=1\times 10^{-7}M_\odot/yr$ disc with and without a $\alpha$ transition.
   \label{fig:SurfHrdead}
   }
\end{figure}

In Fig.~\ref{fig:SurfHrdead} (top) the midplane temperature profiles for $\dot{M}=1\times 10^{-7}M_\odot/yr$ discs with and without a transition in $\alpha$ are shown. Up to $\approx 20$AU the midplane temperature is smaller in the case with $\alpha$ transition than the case without $\alpha$ transition. Consequently, the $H/r$ profile of the disc with $\alpha$ transition is smaller than the $H/r$ profile of the disc without $\alpha$ transition (second panel in Fig.~\ref{fig:SurfHrdead}). Here we define $H$ as the pressure scale height using the midplane sound speed $c_s (z=0)$ with $H=c_s(z=0) / \Omega_K$. The bumps in the $H/r$ and $T$ profiles at $\approx 4$AU are also caused by transitions in the opacity profile (see Paper II), which changes the heating and cooling properties of the disc.

\begin{figure}
 \centering
 \includegraphics[scale=0.70]{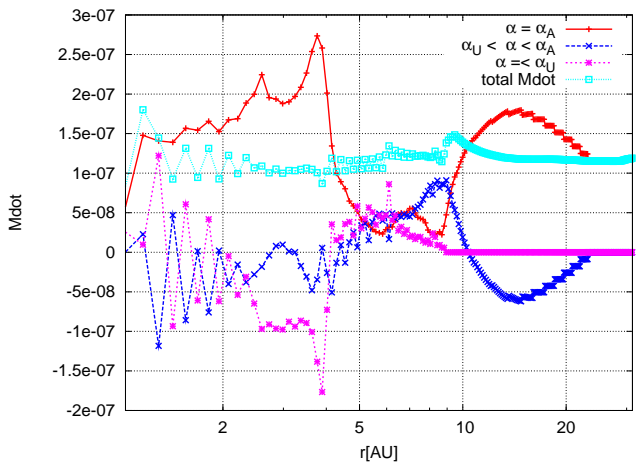}
 \caption{$\dot{M}$ rates for the different layers of the $\dot{M}=1\times 10^{-7}M_\odot/yr$ disc with an $\alpha$ transition. Here, a positive $\dot{M}$ rate indicates accretion, while a negative $\dot{M}$ indicates decretion (motion away from the star). 
   \label{fig:flow107}
   }
\end{figure}

In a 1D radial model of a disc with constant radial $\dot{M}$, a reduction of $\alpha$ is compensated for by an increase of $\Sigma_G$ of the same magnitude, as $\Sigma_G \propto \dot{M} / \nu$. This would allow for the same amount of viscous heating in the 1D disc and hence the same temperature profile. However, in $r-z$-discs this is not the case. A reduction in $\alpha$ does not imply an increase of $\Sigma_G$ by the same factor, because the active layer of the disc can carry most of the accretion rate (see Fig.~\ref{fig:flow107}), which is not possible in 1D models. Indeed, the surface density profile in the inner part of the disc with $\alpha$ transition (third from top in Fig.~\ref{fig:SurfHrdead}) only shows an increase by a factor of $3$ in contrast with the reduction in $\alpha$ by a factor of $100$ in the dead zone.

In the outer parts of the disc, the surface density is lower for the disc with an $\alpha$ transition. This is caused by the higher $H/r$ in the outer parts, which results in a higher viscosity and hence in a lower surface density for an equal $\dot{M}$ disc.

In the disc with an $\alpha$ transition, the surface density shows a significant local maximum at $\approx 4$AU, which is exactly where we set the transition of $\alpha_U$ to $\alpha_D$. This is because, as $\dot{M}$ is independent of $r$, a change in viscosity is compensated for by a change in $\Sigma_G$, resulting in the high-density regions close to the location of the change in $\alpha$. This local maximum in surface density at $\approx 4$AU is very appealing, because it might function as a planet trap (by changing the disc vortensity) and stop inward migration of low-mass planets that migrate in the type-I-migration regime \citep{2006ApJ...642..478M}. In contrast, the disc without an $\alpha$ transition shows a dip in surface density around $\approx 4$AU, which is caused by the transition in opacity at the ice line and the corresponding increase in viscosity caused by the increase in $H/r$ (see Paper II). 

The midplane pressure of the same discs is displayed in the bottom panel of Fig.~\ref{fig:SurfHrdead}. For the disc without an $\alpha$ transition, the pressure follows a power law without any derivations. No pressure bumps exist in this case. In the case with an $\alpha$ transition, however, small wiggles in the pressure are visible, especially at $\approx 4$AU, where $\alpha_U$ transitions to $\alpha_D$. 

Despite the bumps in surface density, no real pressure bump exists in the disc because the gradient of pressure is not inverted. In 2D discs the pressure is defined as 
\begin{equation}
 P_{2D}= R_{gas} \Sigma_G T / \mu \propto \Sigma_0 T_0 r^{-s-\beta} \ , 
\end{equation}
with $R_{gas}$ being the gas constant and $\mu$ the mean molecular weight. Here $s$ denotes the power-law index of the surface density profile and $\beta$ the power-law index of the temperature profile. This indicates that when a jump in surface density can compensate for the gradient in $T$, a pressure bump exists. However, in 3D discs the pressure is related to the volume density $\rho_G$ as
\begin{equation}
  \label{eq:pressure}
  P_{3D} = R_{gas} \rho_G T / \mu \propto \frac{\Sigma_0}{H_0} T_0 r^{-s-\beta-(1+f)} \ ,
\end{equation}
as in hydrostatic equilibrium $\rho_G = \Sigma_G / (\sqrt{2 \pi} H)$, where $H$ is the disc vertical thickness. Here $f$ denotes the flaring index of the disc. By comparing $P_{3D}$ with $P_{2D}$, we immediately see that for the same temperature profile a steeper positive radial surface density gradient is needed in 3D than in 2D to invert the radial pressure gradient, as long as the flaring index of the disc is not strongly negative, which generally is the case. In the models shown in Fig.~\ref{fig:SurfHrdead}, the local maximum of the surface density gradient is not large enough to result in a pressure bump.

\begin{figure}
 \centering
 \includegraphics[scale=0.78]{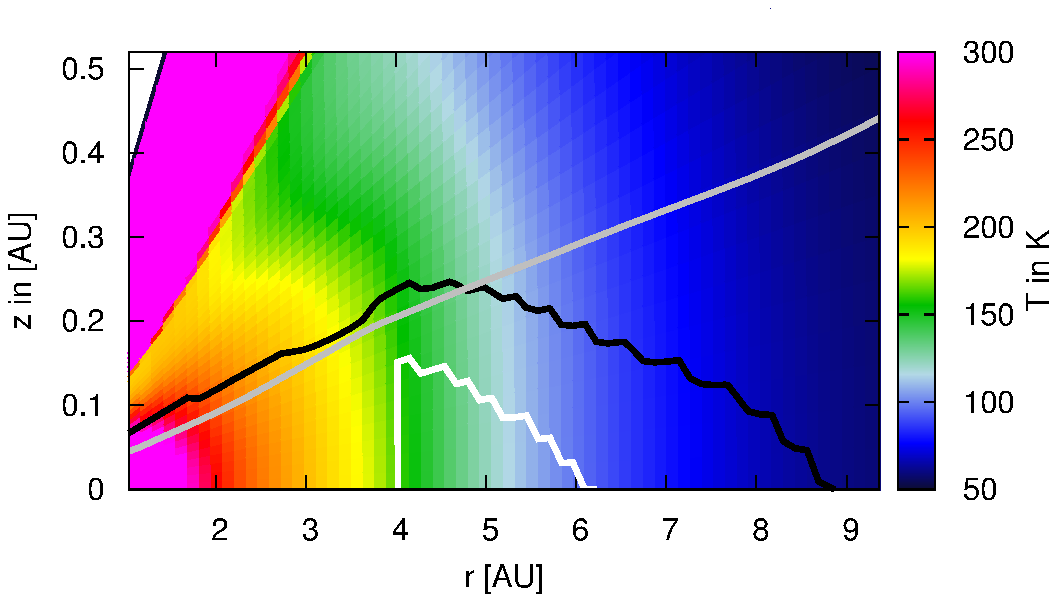}
 \includegraphics[scale=0.78]{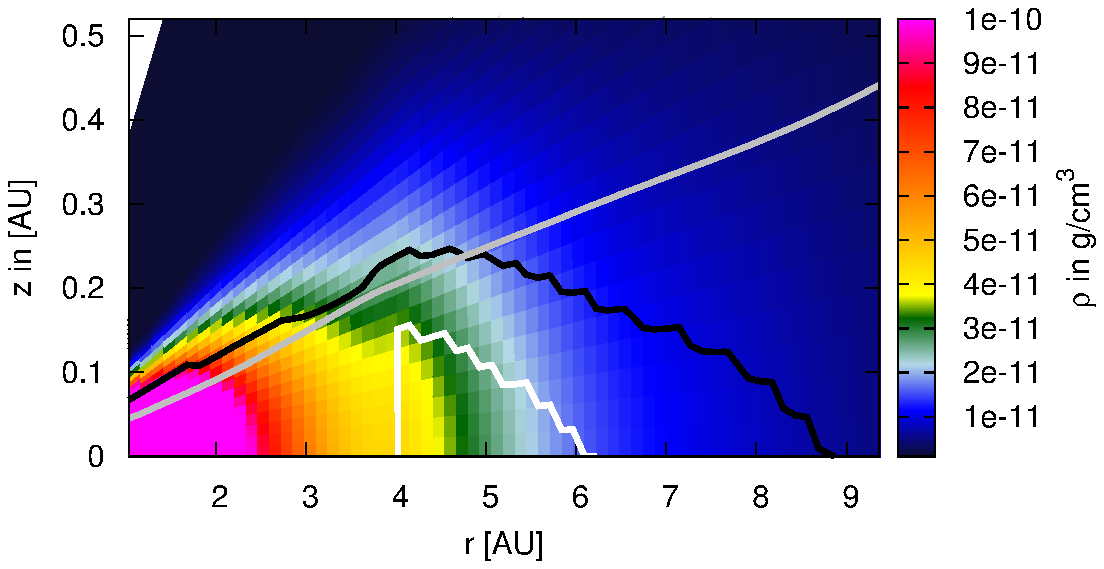}
 \caption{Temperature (top) and density (bottom) of the $\dot{M}=1\times 10^{-7}M_\odot/yr$ with $\alpha$ transition. The black and white lines mark the transitions in the $\alpha$ viscosity parameter as defined by eq.~\ref{eq:alpha} and as shown in Fig.~\ref{fig:Alpha}. The grey line marks the disc's {\bf pressure scale} height $H$.
   \label{fig:RhoT2D}
   }
\end{figure}

In Fig.~\ref{fig:RhoT2D} the ($r,z$) temperature (top) and density (bottom) distribution are shown. In the inner parts close to midplane, the disc is heated by viscosity. With increasing distance to the star, this heating decreases as the density decreases. The upper layers of the disc are heated by the star, therefore they are hotter than the midplane region. The regions of the disc far away from the star with low density are heated by heat diffusion downwards from the upper layers. Close to the $\alpha$ transition at $\approx 4$AU, the temperature is vertically constant for a very wide vertical range (up to $\approx 0.3$AU).

The volume density $\rho_G$ near the midplane is nearly radially constant around the $\alpha$ transition at $\approx 4$AU, as shown in Fig.~\ref{fig:RhoT2D} (bottom). In this region  the density is vertically constant up to $\approx 0.1$AU. To calculate the surface density, the density shown here was vertically integrated, which resulted in the local maximum in the surface density (middle panel in Fig.~\ref{fig:SurfHrdead}). But, this is not enough to generate a pressure bump because $T$ decreases with $r$.

\begin{figure}
 \centering
 \includegraphics[scale=0.78]{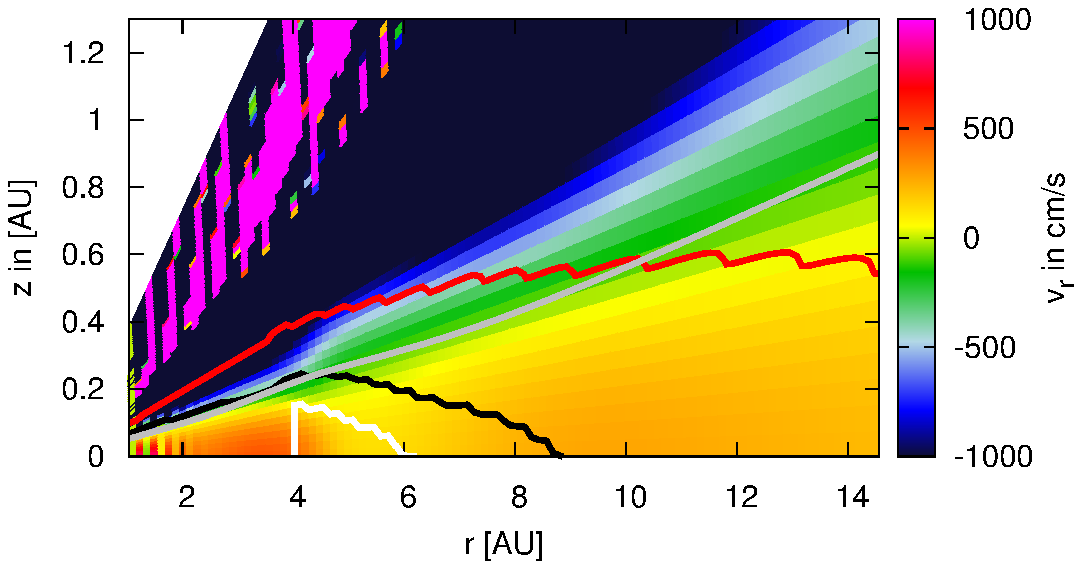}
 \includegraphics[scale=0.78]{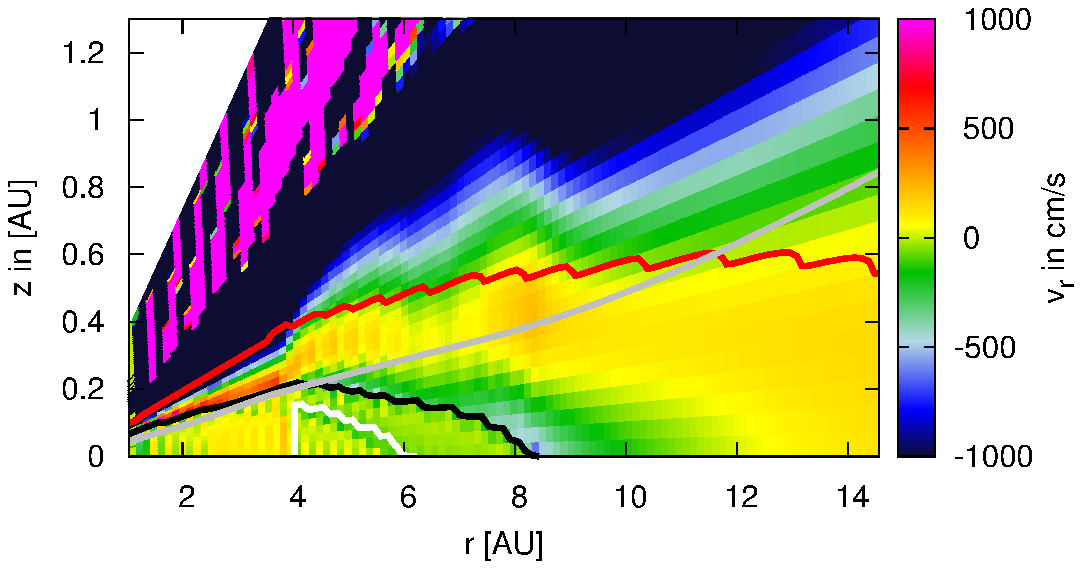}
 \caption{Radial velocity of the disc without $\alpha$ transition (top) and with $\alpha$ transition (bottom). The black and white lines mark the transitions in the $\alpha$ viscosity parameter as defined by eq.~\ref{eq:alpha} and as shown in Fig.~\ref{fig:Alpha}. In the bottom panel, the parts of the disc above the red line are fully active with $\alpha_A$. They are overplotted in the top plot as well to show where these transitions would be. The grey line represents the pressure scale-height $H$. Positive velocities mark an outward motion, while negative velocities indicate an inward motion. 
   \label{fig:velocity107}
   }
\end{figure}

The radial velocity of the disc without an $\alpha$ transition is displayed in the top panel of Fig.~\ref{fig:velocity107}. A positive value indicates an outward motion, while a negative value indicates an inward motion. Here, we see an outflow in the midplane regions of the disc and an inflow at the top layers of the disc, which was also observed by \citet{1984SvA....28...50U, 1992ApJ...397..600K, Takeuchi2002}. This flow is caused by the $\alpha$ prescription for the viscosity \citep{2011A&A...534A.107F}. 

The radial velocity of the disc with $\alpha$-transition is displayed in the bottom panel of Fig.~\ref{fig:velocity107}. We now observe several layers in the flow pattern. In the very top regions of the disc (where $\alpha=\alpha_A$) the flow is directed inwards. Farther down, the flow is directed outwards, while it then flows inwards again at the $\alpha_U$ transition. The flow farther inside the disc near the midplane is again directed outwards. Both discs show similar inflow behaviour in the top layers.

In Fig.~\ref{fig:flow107} the $\dot{M}$ rates for the different $\alpha$-layers of the disc with $\alpha$ transition are displayed. The layers are defined by their $\alpha$ values as indicated in the figure. Additionally, the total $\dot{M}$ rate is displayed, which is the sum of all the different layers. While the total $\dot{M}$ rate remains approximately constant in radius, the different $\dot{M}$ rates of the different layers vary strongly. Between $2-4$AU, the $\dot{M}$ rate in the active layer is highest (in inward direction) and in the same region, the layer with $\alpha<\alpha_U$ has the largest outflow. At $4$AU the flux changes and each layer seems to carry the same amount of $\dot{M}$. At $8$AU, $\alpha > \alpha_U$, so that only two layers in the disc exist. At that distance the flow returns to the state of an unlayered disc, with outflow in midplane and inflow in the top layers. We recall that the accretion rate of the active layer only depends on the radial velocity of the gas, because its density is fixed to $10g/cm^2$ at each $r$ by construction.

In a simple view of a dead zone, with zero viscosity, the material flows inwards in the top layers of the disc and then dumps material into the dead zone, which then grows in density. Eventually, the dead zone becomes so massive that the disc becomes gravitationally unstable and produces outbursts of accretion \citep{2014MNRAS.437..682M}. But this view is not supported by the flow picture we found in our simulations. Just above the region of low viscosity, we found outflow and in the very top layers inflow. This suggests that the mass flux in the active layer can adjust to the dead zone so that the mass flux is still constant in $r$ and at the same time does not dump that much material in the dead zone. This consideration is also supported by the fact that we did not find an increase of $\Sigma_G$ of a factor of $100$ in the dead zone in our simulations, where $\alpha$ is reduced by a factor of $100$. This basically suggests that the discs cannot become gravitationally unstable which disagrees with the hypothesis of \citet{2014MNRAS.437..682M}.

\subsection{Planetesimal formation and embryo migration}
\label{subsec:migration}

The reduction of the gravitational force by the radially outwards-pointing force of the pressure gradient causes a difference $\Delta v$ between the azimuthal mean gas flow and the Keplerian orbit. This is given by
\begin{equation}
\label{eq:stream}
  \frac{\Delta v}{c_s} = \eta \frac{v_K}{c_s} = - \frac{1}{2} \frac{c_s}{v_K} \frac{d\ln (P)}{d\ln (r)} \ ,
\end{equation}
where $v_K=\sqrt{GM/r}$ is the Keplerian velocity $c_s / v_K = H/r$ (for the $H/r$ profile, see the second panel from top in Fig.~\ref{fig:SurfHrdead}). $\eta$ represents a measure of the gas pressure support \citep{1986Icar...67..375N}.

Because of the sub-Keplerian rotation of the gas, small solid particles drift towards the star. A local reduction in $\frac{\Delta v}{c_s}$ facilitates particle clumping in the streaming instability \citep{2007Natur.448.1022J,2007ApJ...662..627J,2010ApJ...722.1437B,2010ApJ...722L.220B}. Reduced $\frac{\Delta v}{c_s}$ would also help planetesimals to grow further by pebble accretion \citep{2012A&A...544A..32L}. 

\begin{figure}
 \centering
 \includegraphics[scale=0.7]{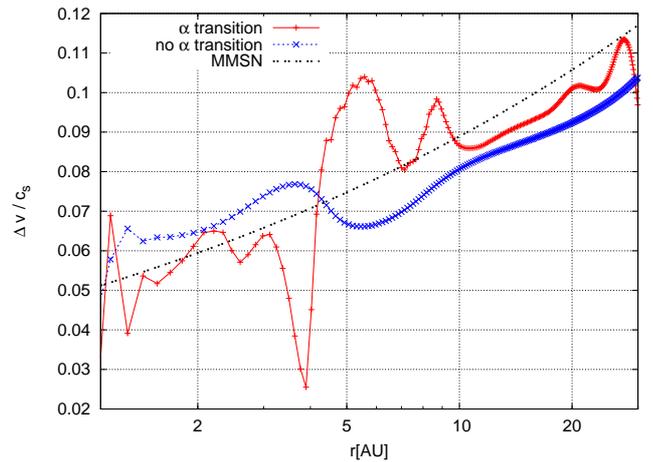}
 \caption{$\frac{\Delta v}{c_s}$ of the $\dot{M}=1\times 10^{-7}M_\odot/yr$ disc with and without an $\alpha$ transition. The black line marks the $\Delta$ parameter in the standard MMSN disc.
   \label{fig:eta107}
   }
\end{figure}

The $\frac{\Delta v}{c_s}$ parameter is displayed in Fig.~\ref{fig:eta107}. In addition to the $\dot{M}=1\times 10^{-7}M_\odot/yr$ discs with and without $\alpha$ transition, we display the $\frac{\Delta v}{c_s}$ parameter in the standard Minimum Mass Solar Nebular (MMSN) case \citep{1981PThPS..70...35H}. The MMSN model is constructed with simple power laws in surface density and temperature, which results in a power law for the $\frac{\Delta v}{c_s}$ parameter as well. The increasing $\frac{\Delta v}{c_s}$ with $r$ means that it is much harder to form planetesimals via the streaming instability at larger distances.

The disc without transition in $\alpha$ has a maximum at $r \approx 3.5$ AU and a drop for larger $r$ that is caused by the maximum in the $H/r$ profile, which is originally caused by transitions in opacity (Paper II). That $\frac{\Delta v}{c_s}$ has the same value at $2$AU and at $5$AU implies that the formation of planetesimals at these locations are equally likely. But at $5$AU the disc is much colder than at $2$AU, which allows for the condensation of ice grains, meaning that more solid material is available at $5$AU. In the end, this might actually facilitate planetesimal formation more at $5$AU than at $2$AU.

In the case with an $\alpha$ transition $\frac{\Delta v}{c_s}$ is much reduced at $4$AU. This is the exact location of the $\alpha$ transition from $\alpha_U$ to $\alpha_D$. This transition causes a wiggle in the midplane pressure (bottom panel in Fig.~\ref{fig:SurfHrdead}). The change of the pressure gradient then significantly reduces $\frac{\Delta v}{c_s}$ at that location, making planetesimals formation via the streaming instability much more likely there.

After they are formed, small planetesimal can migrate inwards. These planetesimals can then grow to planetary embryos by oligarchic growth \citep{1998Icar..131..171K} or by pebble accretion \citep{2012A&A...544A..32L, 2012A&A...546A..18M}. We now shift our attention to planet migration. To measure the migration rate, one should in principle extend our disc model to full 3D (with sufficient resolution in azimuth), introduce a planet in the disc and measure the torques. This would be very computationally expensive, however, in particular because for the small-mass planets that we are interested in here, a very high resolution would be required to resolve the planet's horseshoe region. To estimate the smallest mass needed for outward migration of embryos triggered by the entropy related corotation torque, we therefore used the formula by \citet{2011MNRAS.410..293P}, which captures the effects of torque saturation due to viscosity effects in contrast to \citet{2010MNRAS.401.1950P}, where the torques are fully unsaturated. The torque formula by \citet{2011MNRAS.410..293P} was observed to match with the torques in 3D simulations quite well \citep{2011A&A...536A..77B}. However for small-mass planets, a new phenomenon has been observed that can result in a more negative torque that drives a faster inward migration than described by the formula of \citet{2011MNRAS.410..293P} \citep{Lega2013}. However, this phenomenon deserves more detailed studies. We therefore used the formula by \citet{2011MNRAS.410..293P} as an estimate for planet migration in gas discs. This was made in the same way as described in Papers I and II, so we do not state the torque formula explicitly. The total torque acting on an embedded planet is a composition of its Lindblad torque and its corotation torque:
\begin{equation}
 \Gamma_{tot} = \Gamma_L + \Gamma_C \ .
\end{equation}
The Lindblad and corotation torque depend on the local radial gradients of entropy $S \propto r^{-\xi}$, with $\xi = \beta - (\gamma - 1.0) s$, and gas surface density $\Sigma_G \propto r^{-s}$. $\beta$ describes the radial gradient of the temperature profile, $T \propto r^{-\beta}$. Very approximately, for large $s$ a large $\xi$ caused by a large $\beta$ will lead to outward migration, while a flat radial entropy gradient will lead to inward migration.

\begin{figure}
 \centering
 \includegraphics[scale=0.78]{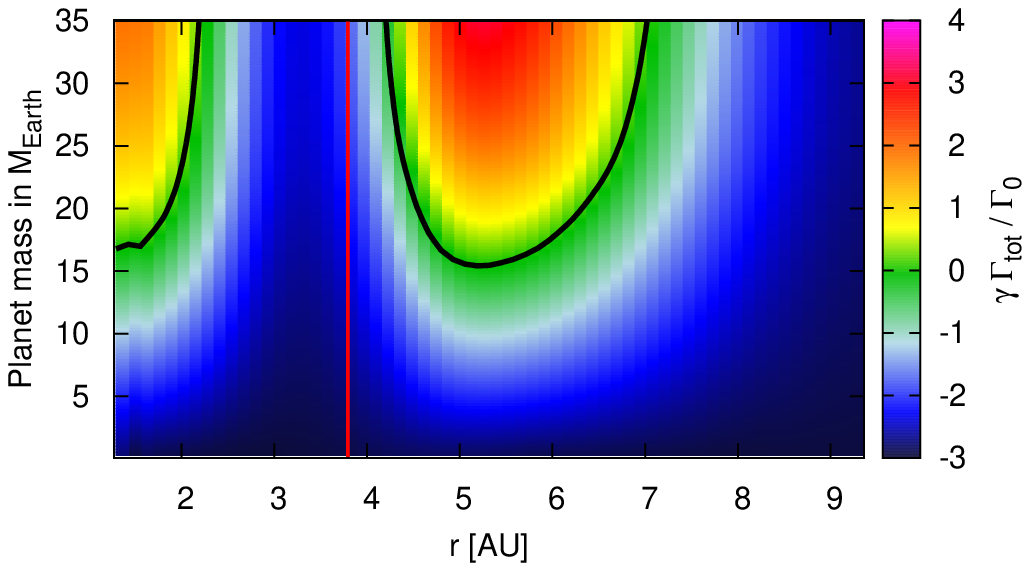}
 \includegraphics[scale=0.78]{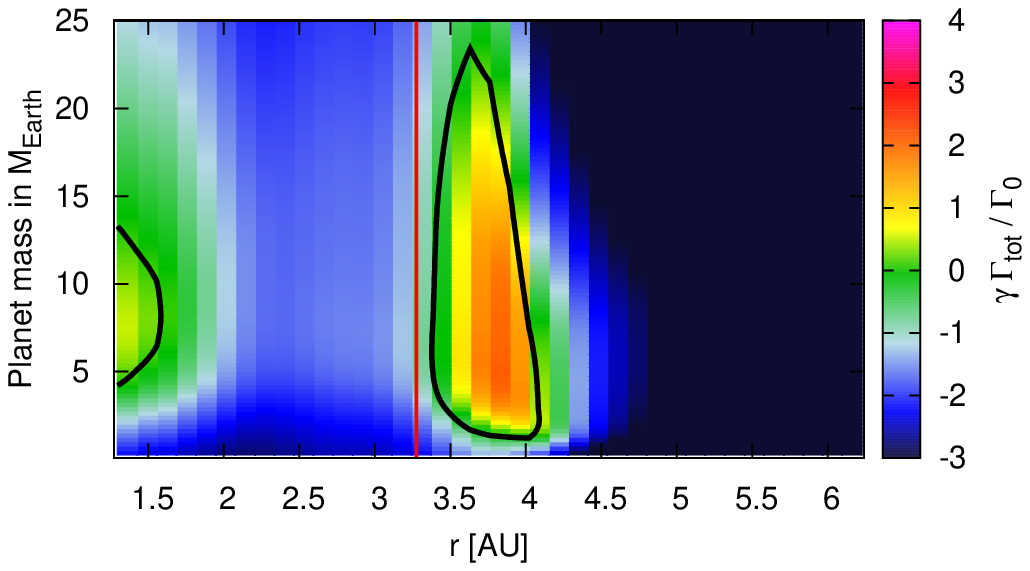}
 \caption{Migration map for discs without (top) and with (bottom) an $\alpha$ transition. The encircled regions in black mark regions in which outward migration is possible. The vertical red line marks the ice line at $190$K in both plots. The two plots feature slightly different radial extensions and planetary masses.
   \label{fig:Mdot107dead}
   }
\end{figure}

The migration maps for discs without (top) and with (bottom) an $\alpha$ transition are displayed in Fig.~\ref{fig:Mdot107dead}. In the case without an $\alpha$ transition, two regions of outward migration are visible, one around $2$AU and one between $4$ and $7$AU. For these regions of outward migration a planetary mass of at least $\approx 17 M_{Earth}$ is needed, which is very high. The reason for this lies in the relatively high viscosity with $\alpha=0.0264$ (Paper II).

In the case of an $\alpha$ transition this changes dramatically. There are still two regions of outward migration, but they are much more narrow in radial distance than for the case without an $\alpha$ transition. 

More precisely, now the region of outward migration is only a small band between $3.5$ to $4.0$AU. This is because farther away the viscosity is too low to keep the entropy-driven corotation torque unsaturated. The outer migration region is now due to the vortensity-driven corotation toque, which is positive and locally exceeds the Lindblad torque because of the positive radial gradient of the surface density. This is the planet-trap mechanism first studied in \citet{2006ApJ...642..478M}. \footnote{In a disc without an $\alpha$ transition the outer radial boundary of the outward migration region is approximately the location where the aspect ratio of the disc has a minimum, i.e. at the transition between the part of the disc that is dominated by viscous heating and that dominated by stellar irradiation. Thus, changing the luminosity properties of the central star affects the extension of the outward migration region. In a disc with an $\alpha$ transition the outward migration region is instead located near the maximum of $H/r$, and thus it is entirely in the part of the disc dominated by viscous heating. Therefore, the outward migration region becomes quite insensitive to the assumed stellar properties.} The planet trap mechanism also changes the smallest and largest masses for outward migration, relative to the entropy-driven corotation case. The smallest mass for outward migration is now $\approx 0.5M_{Earth}$, which is much smaller than before. On the other hand, the largest mass for undergoing outward migration is reduced to $\approx 23M_{Earth}$.

Planetesimal formation for discs with and without the $\alpha$ transition, the streaming instability is more likely to operate in a region of outward migration. However, this may not be enough to explain the formation of a giant planet at the snowline (i.e. at the location of the $\alpha$-transition). In fact, only objects more massive than $0.5 M_{Earth}$ avoid inward migration in this region. Thus, if the streaming instability and the subsequent pebble accretion \citep{2012A&A...544A..32L} do not form a $0.5 M_{Earth}$ object sufficiently fast, the object will have the time to migrate into the inner system. Therefore, a steeper viscosity gradient than assumed here may be needed to allow a giant planet to form at the snowline to produce a steeper surface density gradient (capable of stopping the inward migration of smaller-mass planets) and possibly even a pressure bump (preventing the inward migration of planetesimals).

\section{Disc evolution}
\label{sec:discevolve}

We follow here the prescription for the transition in $\alpha$ from eq.~\ref{eq:alpha} and decrease the disc's $\dot{M}$ by decreasing the $\Sigma_G$ value to follow the evolution of the disc through different accretional stages (same approach as in Paper II). We expect that with the decrease of the surface density of the disc, the radial gas-flow pattern eventually returns to the pattern of an unlayered disc. This is because the vertical integrated surface density from the disc's surface down to the midplane, which cannot exceed $\Sigma_G/2$, will not exceed the $100$g/cm$^2$ limit to allow a full transition from $\alpha_U$ to $\alpha_D$ (see eq.~\ref{eq:alpha}).

\begin{figure}
 \centering
 \includegraphics[scale=0.7]{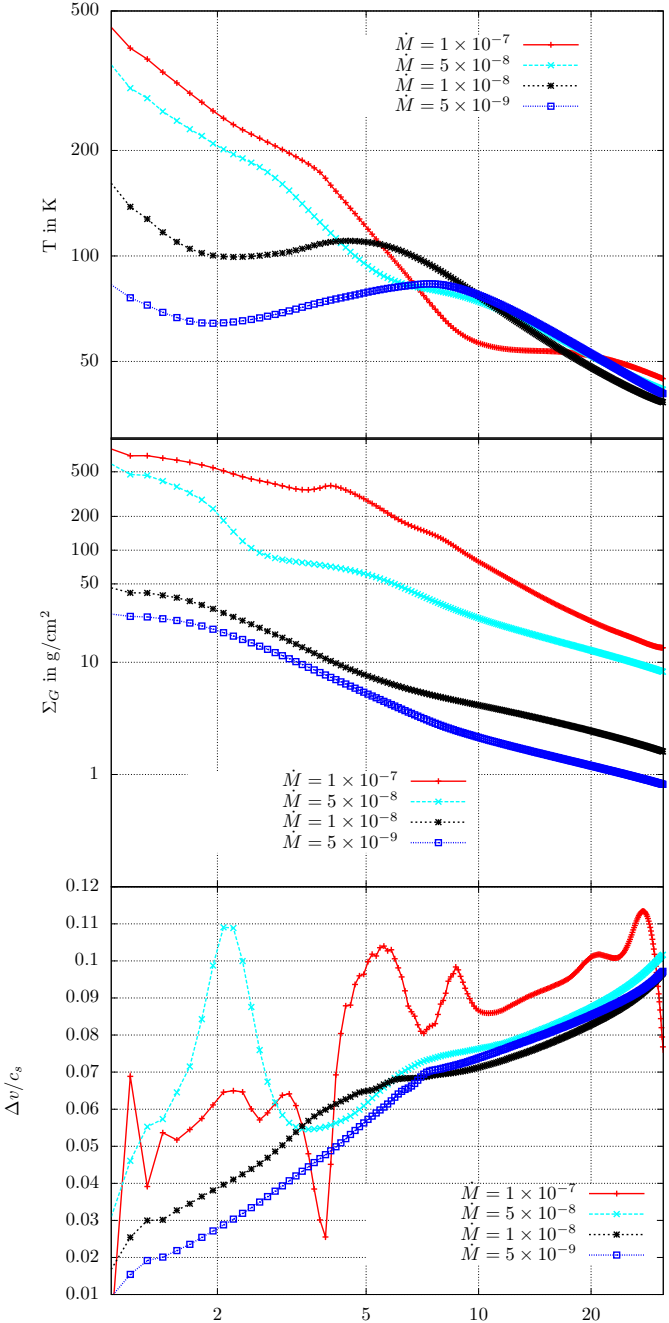}
 \caption{Surface density (top) and midplane temperature (middle) for discs featuring different $\dot{M}$ rates with $\alpha$ transitions. The bottom plot features $\frac{\Delta v}{c_s}$ of the streaming instability for the same discs.
   \label{fig:SigTempall}
   }
\end{figure}

The surface density (top) and midplane temperature (middle) are displayed in Fig.~\ref{fig:SigTempall}. The ratio between the surface densities in the $\dot{M}=1\times 10^{-7}M_\odot/yr$ and the $\dot{M}=1\times 10^{-8}M_\odot/yr$ is higher than a factor of $10$ (the ratio of the accretion rates) in the inner parts of the disc. This is because the disc with the smaller $\dot{M}$ has a larger vertically averaged $\alpha$ (because the column density of the gas is lower). Consequently, the full surface density is reduced more than proportionally to the accretion rate. The outer parts of both discs are fully active, so that the difference in $\Sigma_G$ is $\approx 10$, i.e. proportional to the change of the $\dot{M}$ rate. This means that the change in total surface density $\Sigma_G$ is not linear in the change of $\dot{M}$, because the change of viscosity is not linear in the first place.

A local maximum in the surface density (positive surface density gradient) is visible only for the $\dot{M}=1\times 10^{-7}M_\odot/yr$ disc . For the $\dot{M}=5\times 10^{-8}M_\odot/yr$ disc a dip in the surface density profile at $\approx 2-3$AU is visible, which is at the outer edge of the $\alpha_U$ layer. In fact, inside of $\approx 2.5$AU, this reduction in $\alpha$ causes the surface density to increase. For lower $\dot{M}$ values no such dips are visible, because the highest surface density of these discs at any radius is lower than the condition for reducing $\alpha_A$ fully towards $\alpha_U$.

$\frac{\Delta v}{c_s}$ is displayed in the bottom panel of Fig.~\ref{fig:SigTempall}. The dip in $\frac{\Delta v}{c_s}$ that was visible for the $\dot{M}=1\times 10^{-7}M_\odot/yr$ disc vanishes for smaller $\dot{M}$. This is because the transition in $\alpha$ form $\alpha_U$ to $\alpha_D$ disappears, because the disc's surface density is decreased. The big bump of $\frac{\Delta v}{c_s}$ at $\approx 2$AU in the $\dot{M}=5\times 10^{-8}M_\odot/yr$ disc is caused by the very steep negative surface density gradient in this region, which results in a steeper negative pressure gradient (eq.~\ref{eq:pressure}), which increases $\frac{\Delta v}{c_s}$ (eq.~\ref{eq:stream}). In the later stages of the disc evolution, $\frac{\Delta v}{c_s}$ becomes much smaller in the inner regions of the disc, which in principle facilitates planetesimal formation by the streaming instability. However, at such late stages the lifetime of the disc is quite short, so that there may not be enough time to form the cores and then accrete gas to form giant planets.

\begin{figure}
 \centering
 \includegraphics[scale=0.78]{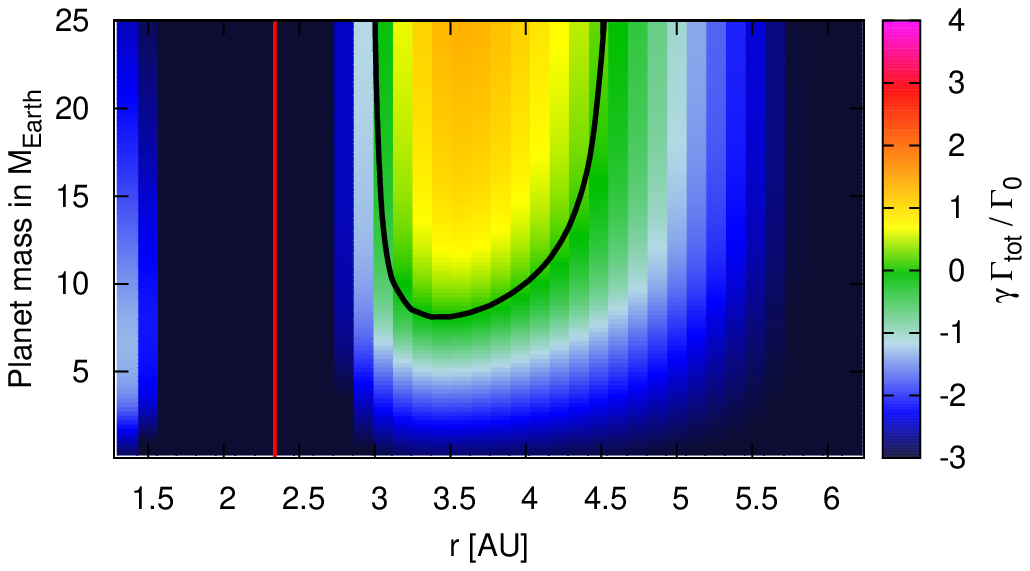}
 \includegraphics[scale=0.78]{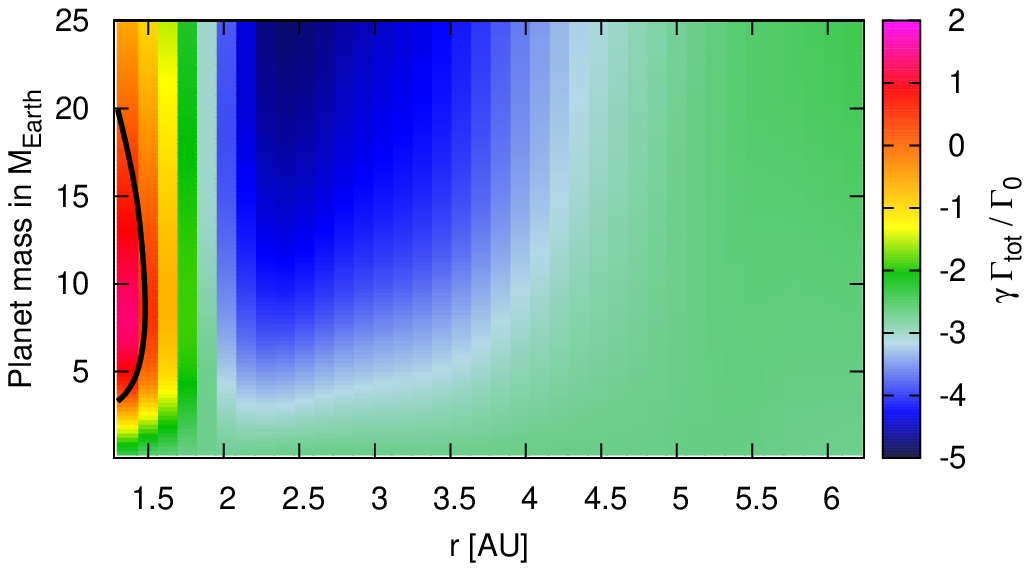}
 \caption{Migration map for discs with an $\alpha$ transition. The discs feature $\dot{M}=5\times 10^{-8}M_\odot/yr$ (top) and $\dot{M}=1\times 10^{-8}M_\odot/yr$ (bottom). The regions encircled in black mark regions in which outward migration is possible. The vertical red line in the top plot marks the ice line at $190$K. The bottom plot has a different colour scale.
   \label{fig:Mdot108dead}
   }
\end{figure}

The migration maps for discs with $\dot{M}=5\times 10^{-8}M_\odot/yr$ (top) and $\dot{M}=1\times 10^{-8}M_\odot/yr$ (bottom) discs are displayed in Fig.~\ref{fig:Mdot108dead}. A large region of outward migration exists around $\approx 3-4.5$AU in the $\dot{M}=5\times 10^{-8}M_\odot/yr$ disc. This region is quite similar to the region of outward migration for the $\dot{M}=1\times 10^{-7}M_\odot/yr$ disc without an $\alpha$ transition shown in the top panel of Fig.~\ref{fig:Mdot107dead}. This is because the disc with $\dot{M}=5\times 10^{-8}M_\odot/yr$ has a surface density lower than $100g/cm^2$ beyond $3$AU and therefore is close to being fully ionised. The region of outward migration here is related to the entropy-driven corotation torque and not to the vortensity-driven corotation torque.

The $\dot{M}=1\times 10^{-8}M_\odot/yr$ disc features for $2<r<4$AU a region of strong inward migration, which is caused by the positive temperature gradient in the disc. In the very inner parts of the disc ($r<1.0$AU) a region of outward migration exists, because of the strong negative temperature gradient, as shown in the top panel of Fig.~\ref{fig:SigTempall}. This migration map is also similar to the migration map of a disc without an $\alpha$ transition. This means that the region of outward migration moves inwards in time as the disc evolves to smaller $\dot{M}$. 

The time the disc spends at high $\dot{M}$ rates ($\dot{M} > 5 \times 10^{-8}M_\odot/yr$) is expected to be very short (a few 100kyr, see \citet{1998ApJ...495..385H}). But in the evolution from $\dot{M} = 1 \times 10^{-8}M_\odot/yr$ to $5 \times 10^{-8}M_\odot/yr$ the lowest mass for outward planet migration increases from $2.5$ to $10$ Earth masses. This means that the formation of a giant planet in this region can occur only if the growth rate of the core is fast enough. Otherwise the core eventually finds itself below the lower mass boundary of the outward migration region and starts to migrate towards the inner disc. Similarly, when the disc reaches $\dot{M}\approx 1 \times 10^{-8}M_\odot/yr$, the outward migration region disappears for all planetary masses. Thus, it is necessary that by this time, the core has reached a large enough mass to accrete a massive envelope and transit to a slow type-II migration mode. We now examine whether such a fast growth is possible. Considering pebble accretion, the growth rates of planetesimals can be estimated (eq.42 \citet{2012A&A...544A..32L})
\begin{equation}
\Delta t_d \approx 8\times 10^6 \left(\frac{\Delta v/ c_s}{0.05}\right)^3 \left( \frac{\rho_p / \rho_G}{0.01} \right)^{-1} \left(\frac{M_0}{10^{-5} M_E}\right)^{-1} \left(\frac{r}{5 AU}\right) yr \ ,
\end{equation}
where $t_d$ defines the growth timescale, $\rho_P$ the pebble density, and $M_0$ the initial seed mass that accretes the pebbles. Assuming that a fraction of $10\%$ of solids is in pebbles and an initial seed mass of $10^{-4}M_E$, the growth time-scale is about $600$kyr, which is similar to the time-scale of the disc evolution according to the observations of \citet{1998ApJ...495..385H}. This indicates that it may be possible for the cores to grow at the same rate as the outward migration region shifts upward in mass while the stellar accretion rate is reduced (see top panel in Fig.~\ref{fig:Mdot108dead}).

\section{Different $\alpha$ transitions}
\label{sec:difftrans}

We explore here different prescriptions for the change of $\alpha$ as a function of $\Sigma_G$ and $T$.

\subsection{Change of the cosmic ray penetration depth}

The penetration of cosmic and X-rays into the disc is of crucial importance for the drive of the MRI. It is still debated how deep cosmic rays penetrate the disc. We therefore changed the transition from $\alpha_A$ to $\alpha_D$ to model a reduced penetration depth of cosmic and x-rays in the following way:
\begin{eqnarray}
\label{eq:alphacosmic}
 10g/cm^2 < \Sigma_P \leq 30g/cm^2 \quad &\Rightarrow& \quad \alpha_A \to \alpha_U \nonumber \\
 30g/cm^2 < \Sigma_P; 220K<T \leq 800K \quad &\Rightarrow& \quad \alpha_U \nonumber \\
 30g/cm^2 < \Sigma_P; 160K<T \leq 220K \quad &\Rightarrow& \quad \alpha_U \to \alpha_D \nonumber \\
 30g/cm^2 < \Sigma_P \leq 50g/cm^2; T<160K \quad &\Rightarrow& \quad \alpha_U \to \alpha_D \nonumber \\
 50g/cm^2 < \Sigma_P; T<160K \quad &\Rightarrow& \quad \alpha_D \ .
\end{eqnarray}
This implies that $\alpha_D$ can still be reached at lower surface density values, hence it can be reached at lower $\dot{M}$ values. Accordingly, we expect to observe the features related to the presence of a dead zone (surface density bump, planet trap, etc.) for discs with smaller $\dot{M}$ with this prescription of $\alpha$ (eq.~\ref{eq:alphacosmic}) and not with that of eq.~\ref{eq:alpha}.

\begin{figure}
 \centering
 \includegraphics[scale=0.78]{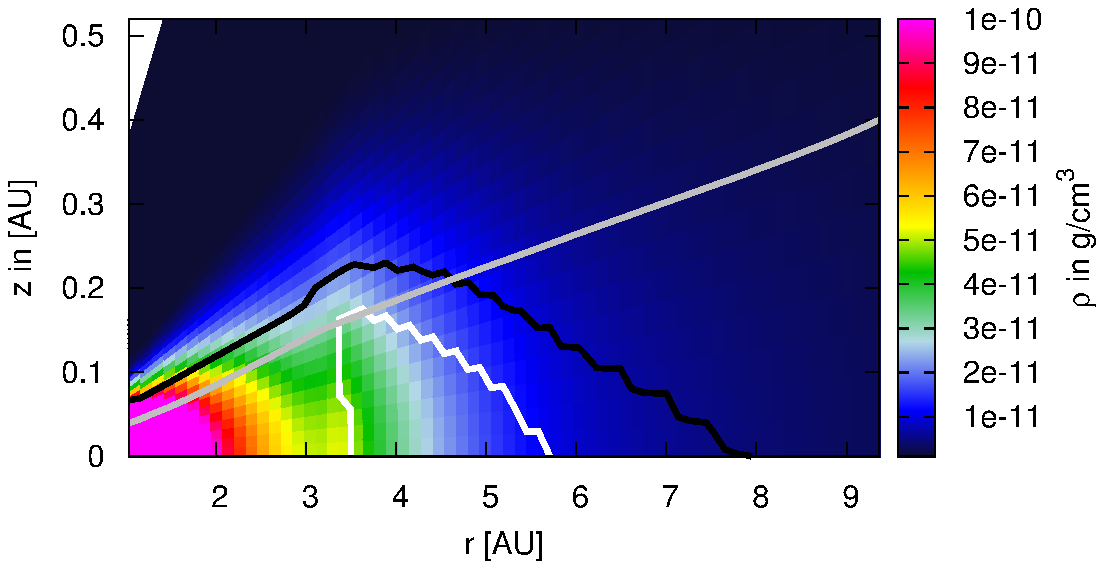}
 \includegraphics[scale=0.78]{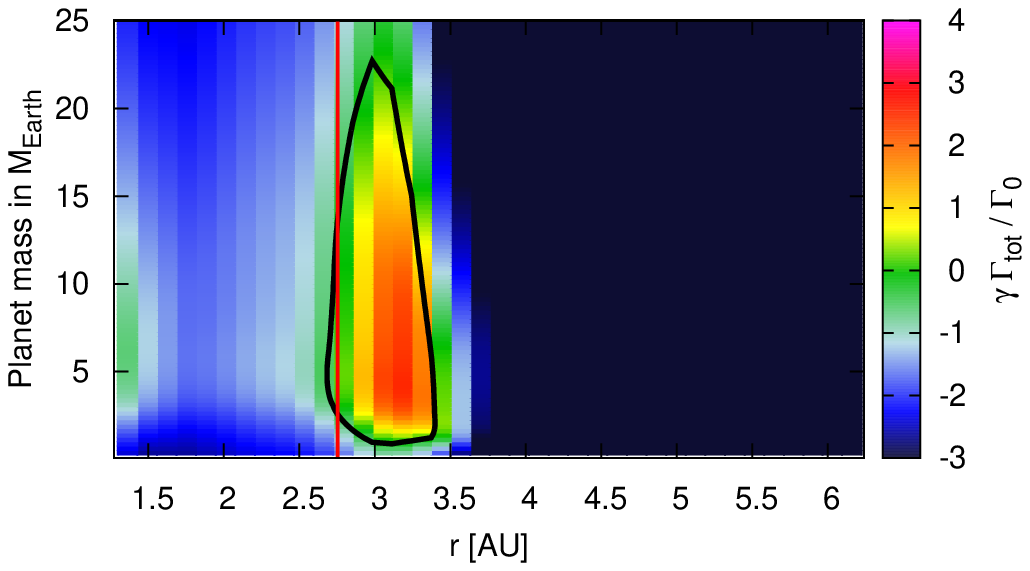}
 \caption{Density (top) and migration map (bottom) of the $\dot{M}=5\times 10^{-8}M_\odot/yr$ disc with a lower penetration depth for cosmic rays. In the top plot, the black and white lines mark the transitions in the $\alpha$ viscosity parameter as defined by eq.~\ref{eq:alphacosmic}, the grey line marks the disc's height $H$. In the bottom plot, the regions encircled in black enclose the regions of outward migration, while the vertical red line marks the ice line at $190$K.
   \label{fig:RhoMigdeadsigma}
   }
\end{figure}

In Fig.~\ref{fig:RhoMigdeadsigma} the ($r,z$) density structure of a $\dot{M}=5\times 10^{-8}M_\odot/yr$ disc with the $\alpha$ transition defined in eq.~\ref{eq:alphacosmic} is shown in the top panel. Clearly, the discs features a local maximum in density at $\approx 3.3$AU. This then translates into a local maximum in the vertically integrated surface density (not displayed). No real pressure bump (positive pressure gradient) is visible in this case either, only a flattening of the pressure gradient, because the negative gradient in $T$ overcompensates for the positive gradient in $\rho_G$ (see eq.~\ref{eq:pressure}).

The migration map (bottom panel in Fig.~\ref{fig:RhoMigdeadsigma}) shows a very similar pattern as that of the $\dot{M}=1\times 10^{-7}M_\odot/yr$ disc with the $\alpha$ transition described by eq.~\ref{eq:alpha} (Fig.~\ref{fig:Mdot107dead}). More precisely, the stopping point of inward migration is shifted slightly inwards to $\approx3$AU. This is expected because the surface density gradients, the $\alpha$ distribution, and the temperature gradients are similar. For smaller $\dot{M}$, the prescription of eq.~\ref{eq:alphacosmic} is not enough to create a low viscosity region. This means that low $\dot{M}$ discs are always nearly fully ionised, and their structure is similar to that described above.

\subsection{Change of the value of $\alpha_U$}

In this section the ratio between the different $\alpha$ is changed, that is, we modify eq.~\ref{eq:alpharatio}. These three independent modifications are
\begin{eqnarray}
 \alpha_A &=& 20 \alpha_U = 100 \alpha_D \ , \nonumber \\
 \alpha_A &=& 50 \alpha_U = 100 \alpha_D \ , \nonumber \\
 \alpha_A &=& 100 \alpha_U = 100 \alpha_D \ ,
\end{eqnarray}
where we kept the transition parameters for temperature and surface density as in eq.~\ref{eq:alpha}. This means that the change of viscosity at the ice line is reduced, because we aim to keep the same total reduction of $\alpha$ from $\alpha_A$ to $\alpha_D$. For the last test case, there is only one transition in $\alpha$ because the transition $\alpha_U \to \alpha_D$ does not exist any more.

\begin{figure}
 \centering
 \includegraphics[scale=0.75]{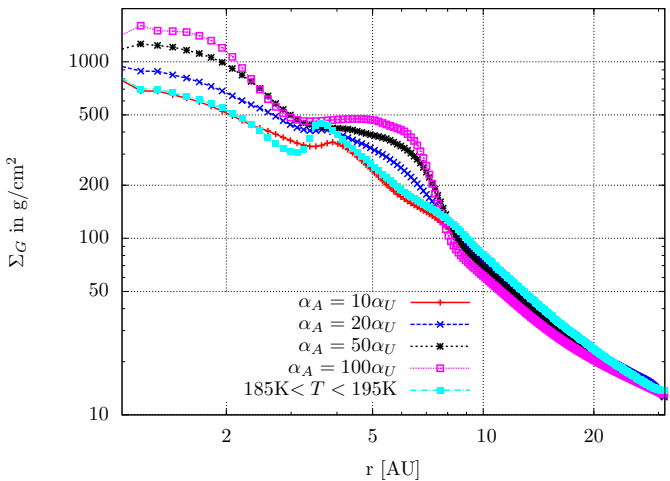}
 \caption{Surface density for $\dot{M} = 1 \times 10^{-7} M_\odot /yr$ discs with different ratios of $\alpha_A/\alpha_U$ (red, blue, and black) and one (light blue) with $\alpha_U \to \alpha_D$ for $185$K$<T<195$K.
   \label{fig:Surfdead2050}
   }
\end{figure}

In Fig.~\ref{fig:Surfdead2050}, the surface density for the four different $\alpha$ ratios for a disc with $\dot{M} = 1 \times 10^{-7} M_\odot /yr$ is displayed. In the outer parts of the disc, the surface density remains the same for all $\alpha$ cases because the outer parts of the disc are fully ionised and hence always have $\alpha_A$, which is the same in all simulations. In the inner parts of the disc, the surface density increases as $\alpha_U$ becomes smaller, because all discs have the same $\dot{M}$.

The local maximum in surface density (positive surface density gradient) at the snow line also deceases when $\alpha_U$ decreases, because the contrast $\alpha_U$ to $\alpha_D$ decreases. In the case of $\alpha_A=50\alpha_U$ the positive surface density gradient at the ice line has completely vanished and only a flat plateau is visible. This might be enough to stop inward migration.

\begin{figure}
 \centering
 \includegraphics[scale=0.78]{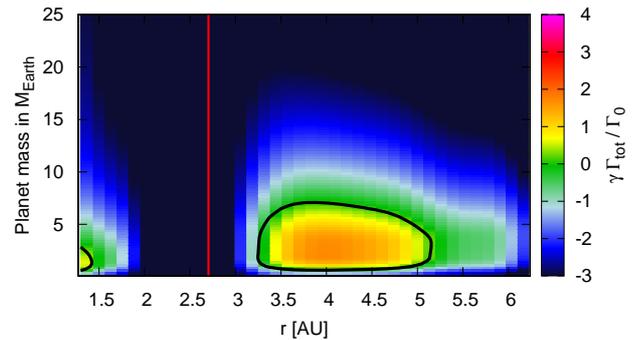}
 \caption{Migration map for the disc with $\dot{M}=1\times 10^{-7}M_\odot/yr$ that features $\alpha_A=100\alpha_U$. The regions encircled in black mark regions in which outward migration is possible. The vertical red line in the plot marks the ice line at $190$K.
   \label{fig:Mdot107dead100}
   }
\end{figure}

The migration map for $\alpha_A=10\alpha_U$ is discussed in section~\ref{sec:visctrans}. We do not display the migration maps for $\alpha_A=20\alpha_U$ and $\alpha_A=50\alpha_U$ here because the migration map looks quite similar to that presented in Fig.~\ref{fig:Mdot107dead} (bottom). The only difference is that the region of outward migration is smaller (e.g. outward migration is still possible only between $3$ and $10M_{Earth}$). In Fig.~\ref{fig:Mdot107dead100} the migration map for $\alpha_A=100\alpha_U$ is displayed. Interestingly, this migration map still features a small region of outward migration, even though that there is only a plateau in the surface density and not a bump.

This is because the barotropic parts of the corotation torque still generate a positive (although weak) contribution, and the steep negative radial temperature gradient in the region produces a positive entropy related corotation torque. The viscosity, although low, is still high enough to prevent torque saturation (this is not the case in the example discussed in Appendix~\ref{ap:interchange}).

\subsection{Change of the transition $\alpha_U$ to $\alpha_D$}

The transition from $\alpha_U$ to $\alpha_D$ in eq.~\ref{eq:alpha} was implemented across a range of $60$K, because this corresponds to the temperature at which the opacity transition is smoothed at the ice line. However, $60$K can be considered a broad temperature range just for the sublimation and condensation of ice grains. We now change this range to $10$K. This is the only modification to eq.~\ref{eq:alpha}. The corresponding surface density is shown in Fig.~\ref{fig:Surfdead2050} as the light-blue line.

\begin{figure}
 \centering
 \includegraphics[scale=0.78]{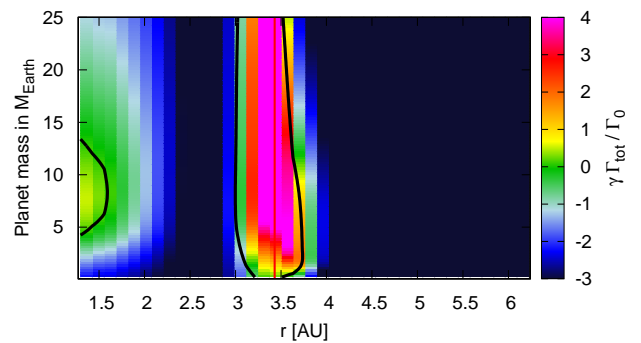}
 \caption{Migration map for the disc with $\dot{M}=1\times 10^{-7}M_\odot/yr$ with steep gradient in the $\alpha$ transition. The encircled regions in black mark regions in which outward migration is possible. The vertical red line in the plot marks the ice line at $190$K.
   \label{fig:Mdot107deadtaper}
   }
\end{figure}

\begin{figure}
 \centering
 \includegraphics[scale=0.75]{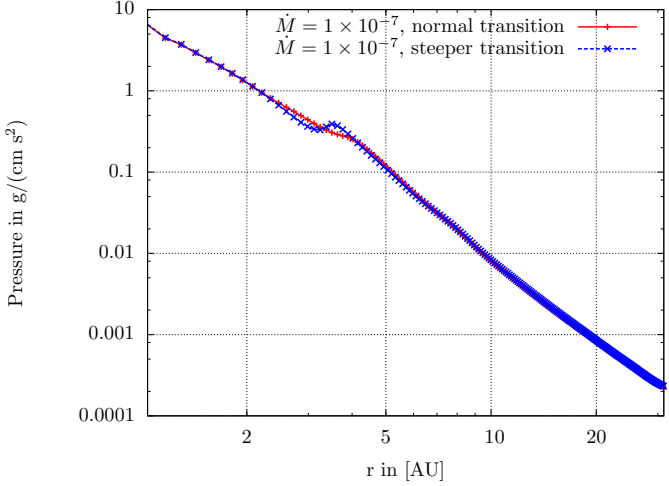}
 \caption{Midplane pressure for the $\dot{M} = 1 \times 10^{-7} M_\odot /yr$ discs with different transition ranges in the temperature for $\alpha_U$ to $\alpha_D$.
   \label{fig:Pressuretaper}
   }
\end{figure}

A much steeper surface density gradient at the location of the ice line is clearly visible. This change of surface density will result in a planet trap, similar to Fig.~\ref{fig:Mdot107dead}, but with a higher intensity. This also traps planets with masses smaller than $0.5M_{Earth}$ (Fig.~\ref{fig:Mdot107deadtaper}). Note that an inverse pressure gradient now exists (Fig.~\ref{fig:Pressuretaper}). An inverse pressure gradient in a protoplanetary disc stops the inward migration of small bodies that undergo gas drag \citep{2008A&A...487L...1B}, promotes the onset of the streaming instability \citep{2007ApJ...662..627J}, and significantly facilitates planetesimals formation \citep{2010ApJ...722.1437B, 2010ApJ...722L.220B}, as also discussed in section~\ref{subsec:migration}.

The maximum steepness a density gradient can have is constrained by the disc height $H$ because a density bump would be unstable to the Rayleigh instability if its width were smaller than $H$ \citep{2010MNRAS.402.2436Y}. Here the transition occurs on a length scale of $2H$.

\section{Gradients at the $\alpha$ transition}
\label{sec:presgrad}

We have presented here detailed simulations of discs with transitions in the $\alpha$ parameter that resulted in changes in the disc structure. If the $\alpha$ transition is large enough, small mass planets can be trapped. The smallest mass of trapped planets decreases with increasing value of the radial positive surface density gradient. At the same location of the disc, $\frac{\Delta v}{c_s}$ of the streaming instability (eq.~\ref{eq:stream}) is greatly reduced, which facilitates planetesimal formation. Again, an inversion of the pressure gradient is only visible if the disc features a very steep positive surface density gradient that can overcompensate for the negative temperature gradient (eq.~\ref{eq:pressure}). 

\begin{figure}
 \centering
 \includegraphics[scale=0.7]{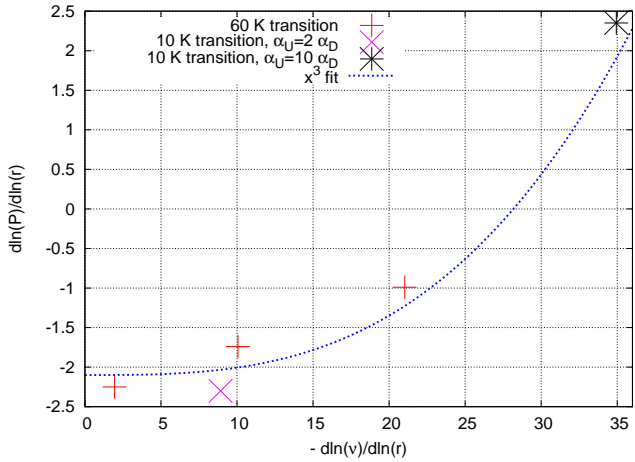}
 \caption{Pressure gradient at the $\alpha$ transition as a function of the viscosity gradient. The data are taken from simulations of the discs with $\dot{M}=1\times 10^{-7}M_\odot/yr$ and different $\alpha$ transitions.
   \label{fig:Pnu}
   }
\end{figure}

We now investigate which changes in $\alpha$, or more precisely, in viscosity are needed to create a pressure bump in the disc. In Fig.~\ref{fig:Pnu} the pressure gradient, $d\ln(P)/d\ln(r)$, is displayed as a function of the viscosity gradient, $d\ln(\nu)/d\ln(r)$. In addition to the simulation data, we provide a rough fit through the data points to estimate the viscosity gradient necessary to cause a pressure bump in the disc.

The change in viscosity required to observe a pressure bump is quite large, $d\ln(\nu)/d\ln(r)\approx -28$. The reduction in viscosity needs to be this high because of the vertical structure of the disc. As stated above, in a 1D (radial) disc, a reduction in viscosity will be equally compensated for by an increase in surface density. If the vertical structure is taken into account, however, this is different because most of the accretion rate can be carried by the active layer (see Fig.~\ref{fig:flow107}), which results in a lower density increase in the midplane regions of the disc. Clearly, a pressure bump is much harder to achieve in a disc with a layered vertical structure than in a 1D disc structure. 

The only simulation in which we observed a bump in the midplane pressure featured a jump of $\alpha$ by a factor of $10$ over a temperature range of $10$K, which corresponds to $0.13$AU in this simulation. We justified this jump in $\alpha$ by more particles being available beyond the ice line ($r>r_{ice}$) that can quench the MRI \citep{2007ApJ...664L..55K}. It seems realistic that the temperature range over which water vapour condensates into ice grains is only $10$K. But a change of $\alpha$ of a factor of $10$ caused by more ice grains is doubtful. A smaller reduction of $\alpha$ seems more reasonable. However, if $\alpha$ is only changed by a factor of $2$, no bump in the midplane pressure can be observed. This indicates the reduction in $\alpha$ needed to create a pressure bump is $\approx 5$.

\section{Summary}
\label{sec:summary}

We have investigated the influence of viscosity transitions on the structure and migration rate of planets in accretion discs, which feature the same mass flow $\dot{M}$ through every radial section of the disc. The structure of the disc was calculated by using 2D hydrodynamical simulations that feature viscous and stellar heating as well as radiative cooling. The viscosity transitions were implemented by reducing the $\alpha$ parameter of the viscosity prescription. When the viscosity is reduced in parts of the disc, the disc's flow adapts to keep the vertically integrated $\dot{M}$ independent of radius.

For high $\dot{M}$ rates, the gas surface density of the disc $\Sigma_G$ is high enough to shield the inner parts of the disc from cosmic and X-rays, resulting in a region of reduced $\alpha$, the dead zone. This region of reduced viscosity is denser than the active layers, which creates a local maximum in surface density. Because we imposed a radial $\alpha$ transition at the ice line as well, we created a planet trap at $\approx 4$AU. This planet trap is very efficient in trapping small-mass planets ($M_P>0.5M_{Earth}$), in contrast to discs without a transition in $\alpha$, where a much larger planetary mass is needed to stop inward migration by the positive corotation torque (Fig.~\ref{fig:Mdot107dead}).

As the disc evolves to lower $\dot{M}$ rates, the surface density decreases, so that the disc is unable to completely shield itself from cosmic and X-rays. This implies that as the disc evolves, the regions of low $\alpha$ shrink and the disc finally reaches a state without reduced $\alpha$. In this case the only possibility to stop planet migration is the positive entropy related corotation torque (when it exists).

The migration of small solids (pebbles) is instead related to gas drag and is sensitive to how sub-Keplerian the gas-disc is. A pressure bump is needed to stop the inward migration of pebbles. To create a pressure bump a positive radial volume density gradient is needed that is strong enough to compensate for the negative temperature gradient. This can only be achieved if the viscosity transition is very steep. More precisely, our simulations show that a reduction of about a factor of $5$ of $\alpha$ is needed over a temperature range of $\approx10$K.

We explained the reduction in $\alpha$ at the ice line by more grains that can block the MRI \citep{2007ApJ...664L..55K}. This view is under debate and the magnitude of this change is not known.

Nevertheless, we can conclude here that quenching the MRI at the ice line by ice grains \citep{2007ApJ...664L..55K} can be an effective trap for Earth-mass planets. This does not necessarily imply a pressure bump in the disc structure, however small dust and ice particles can be trapped there. It implies a significantly reduced pressure gradient, which in turn can significantly facilitate planetesimal formation by the streaming instability \citep{2010ApJ...722.1437B, 2010ApJ...722L.220B}.

\begin{acknowledgements}

B. Bitsch and A. Morbidelli have been partly sponsored through the Helmholtz Alliance {\it Planetary Evolution and Life}. We thank the Agence Nationale pour la Recherche under grant ANR-13-BS05-0003-01 (MOJO). We also thank A. Johansen for useful discussions. In addition we thank an anonymous referee for his/her input. The computations were made done on the “Mesocentre SIGAMM” machine, hosted by the Observatoire de la C\^{o}te d'Azur.

\end{acknowledgements}

\appendix
\section{Interchange between $\Sigma$ and $\nu$}
\label{ap:interchange}

In Paper II we mainly focused on discs that have a ten times higher surface density and a ten times smaller $\alpha$ than the discs presented here, which gives the same $\dot{M}$ rate. Because the penetration of cosmic rays into the disc is dependent on the discs surface density, this is a crucial quantity for developing a low viscosity region through the shielding of cosmic rays. Here, we take a disc with $\dot{M} = 1 \times 10^{-8} M_\odot /yr$, but with the viscosity parameter $\alpha_A=0.003$, which is ten times lower than presented in section~\ref{sec:visctrans}.

\begin{figure}
 \centering
 \includegraphics[scale=0.75]{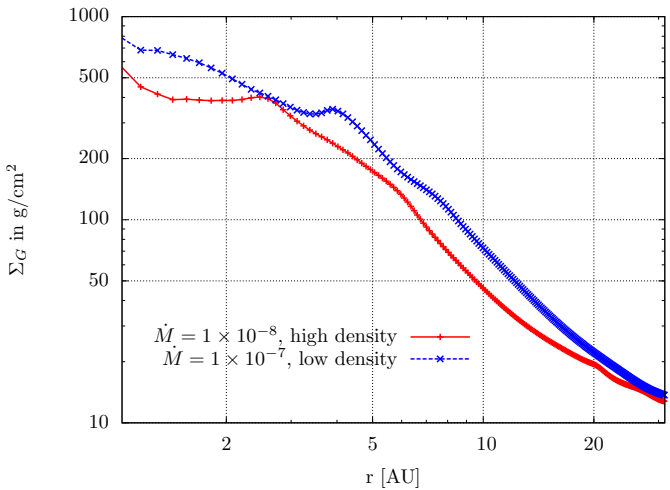}
 \caption{Surface density for $\dot{M} = 1 \times 10^{-8} M_\odot /yr$ disc with $\alpha_A=0.003$, which is $\approx 10$ lower than in Fig.~\ref{fig:SurfHrdead}. The $\dot{M} = 1 \times 10^{-7} M_\odot /yr$ disc with larger $\alpha_A$ is overplotted for comparison.
   \label{fig:Surf108hd}
   }
\end{figure}

In Fig.~\ref{fig:Surf108hd} the surface density of the $\dot{M} = 1 \times 10^{-8} M_\odot /yr$ disc with $\alpha_A=0.003$ is displayed. The prescription of the $\alpha$ transition follows the one given in eq.~\ref{eq:alpha}. As in Fig.~\ref{fig:SurfHrdead} (middle), a bump in the surface density profile is visible. 

\begin{figure}
 \centering
 \includegraphics[scale=0.78]{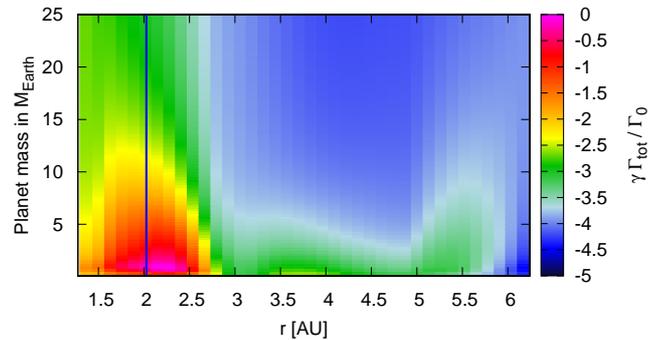}
 \caption{Migration map for $\dot{M} = 1 \times 10^{-8} M_\odot /yr$ disc with $\alpha_A=0.003$, which is $\approx 10$ lower than in Fig.~\ref{fig:Mdot108dead}. The vertical blue line marks the ice line at $T=190$K.
   \label{fig:Mig108hd}
   }
\end{figure}

However, the migration map (Fig.~\ref{fig:Mig108hd}) does not show a positive torque that acts on the planet at any radial distance. The negative torque is smallest in the region close to the surface density bump at $\approx 2.5$AU. This indicates that in principle the bump in the surface density should work as a planet trap, but as the $\alpha$ parameter in that region of the disc is very small ($\alpha_D = 3\times 10^{-5}$), the torques do saturate and cannot sustain outward migration any more (\citet{2006ApJ...642..478M}). This result shows that the parameter range of the $\alpha$ transitions that allow the formation of a planet trap is very narrow.

\bibliographystyle{aa}
\bibliography{Stellar}

\begin{thebibliography}{50}
\expandafter\ifx\csname natexlab\endcsname\relax\def\natexlab#1{#1}\fi

\bibitem[{{Alexander} \& {Pascucci}(2012)}]{2012MNRAS.422L..82A}
{Alexander}, R.~D. \& {Pascucci}, I. 2012, MNRAS, 442, pp.82

\bibitem[{{Bai} \& {Goodman}(2009)}]{2009ApJ...701..737B}
{Bai}, X. \& {Goodman}, J. 2009, ApJ, 701, pp.737

\bibitem[{{Bai} \& {Stone}(2010{\natexlab{a}})}]{2010ApJ...722.1437B}
{Bai}, X.~N. \& {Stone}, J.~M. 2010{\natexlab{a}}, ApJ, 722, pp. 1437

\bibitem[{{Bai} \& {Stone}(2010{\natexlab{b}})}]{2010ApJ...722L.220B}
{Bai}, X.~N. \& {Stone}, J.~M. 2010{\natexlab{b}}, ApJ, 722, L220

\bibitem[{{Balbus} \& {Hawley}(1998)}]{1998RvMP...70....1B}
{Balbus}, S.~A. \& {Hawley}, J.~F. 1998, Reviews of Modern Physics, 70, 1

\bibitem[{{Baruteau} {et~al.}(2014){Baruteau}, {Crida}, {Paardekooper},
  {Masset}, {Guilet}, {Bitsch}, Nelson, {Kley}, \&
  {Papaloizou}}]{2013arXiv1312.4293B}
{Baruteau}, C., {Crida}, A., {Paardekooper}, S.~J., {et~al.} 2014, in
  {Protostars and Planets VI}, arXiv:1312.4293

\bibitem[{{Baruteau} \& {Masset}(2008)}]{2008ApJ...672.1054B}
{Baruteau}, C. \& {Masset}, F. 2008, \apj, 672, 1054

\bibitem[{{Bell} \& {Lin}(1994)}]{1994ApJ...427..987B}
{Bell}, K.~R. \& {Lin}, D.~N.~C. 1994, ApJ, 427, 987

\bibitem[{{Bitsch} {et~al.}(2013){Bitsch}, {Crida}, {Morbidelli}, {Kley}, \&
  {Dobbs-Dixon}}]{2013A&A...549A.124B}
{Bitsch}, B., {Crida}, A., {Morbidelli}, A., {Kley}, W., \& {Dobbs-Dixon}, I.
  2013, A\&A, 549, id.A124

\bibitem[{{Bitsch} \& {Kley}(2011)}]{2011A&A...536A..77B}
{Bitsch}, B. \& {Kley}, W. 2011, A\&A, 536, A77

\bibitem[{{Bitsch} {et~al.}(2014){Bitsch}, {Morbidelli}, {Lega}, \&
  {Crida}}]{2014arXiv1401.1334B}
{Bitsch}, B., {Morbidelli}, A., {Lega}, E., \& {Crida}, A. 2014, astro-ph.EP

\bibitem[{{Brauer} {et~al.}(2008){Brauer}, {Henning}, \&
  {Dullemond}}]{2008A&A...487L...1B}
{Brauer}, F., {Henning}, T., \& {Dullemond}, C.~P. 2008, A\&A, 487, pp.L1

\bibitem[{{Chiang} \& {Goldreich}(1997)}]{1997ApJ...490..368C}
{Chiang}, E.~I. \& {Goldreich}, P. 1997, ApJ, 490, 368

\bibitem[{{Crida} \& {Morbidelli}(2007)}]{2007MNRAS.377.1324C}
{Crida}, A. \& {Morbidelli}, A. 2007, \mnras, 377, 1324

\bibitem[{{Crida} {et~al.}(2006){Crida}, {Morbidelli}, \&
  {Masset}}]{2006Icar..181..587C}
{Crida}, A., {Morbidelli}, A., \& {Masset}, F. 2006, Icarus, 181, 587

\bibitem[{{Dzyurkevich} {et~al.}(2013){Dzyurkevich}, {Turner}, {Henning}, \&
  {Kley}}]{2013ApJ...765..114D}
{Dzyurkevich}, N., {Turner}, N., {Henning}, T., \& {Kley}, W. 2013, ApJ, 765,
  id.114

\bibitem[{{Fromang} {et~al.}(2011){Fromang}, {Lyra}, \&
  {Masset}}]{2011A&A...534A.107F}
{Fromang}, S., {Lyra}, W., \& {Masset}, F. 2011, A\&A, id.A107

\bibitem[{{Hartmann} {et~al.}(1998){Hartmann}, {Calvet}, {Gullbring}, \&
  {D'Alessio}}]{1998ApJ...495..385H}
{Hartmann}, L., {Calvet}, N., {Gullbring}, E., \& {D'Alessio}, P. 1998, ApJ,
  495, p.385

\bibitem[{{Hayashi}(1981)}]{1981PThPS..70...35H}
{Hayashi}, C. 1981, Progress of Theoretical Physics Supplement, 70, pp.35

\bibitem[{{Johansen} {et~al.}(2007){Johansen}, {Oishi}, {Mac Low}, {Klahr},
  {Henning}, \& {Youdin}}]{2007Natur.448.1022J}
{Johansen}, A., {Oishi}, J.~S., {Mac Low}, M.~M., {et~al.} 2007, Nature, 448,
  pp. 1022

\bibitem[{{Johansen} \& {Youdin}(2007)}]{2007ApJ...662..627J}
{Johansen}, A. \& {Youdin}, A. 2007, ApJ, 662, pp. 627

\bibitem[{{Klahr} \& {Bodenheimer}(2003)}]{2003ApJ...582..869K}
{Klahr}, H.~H. \& {Bodenheimer}, P. 2003, \apj, 582, 869

\bibitem[{{Kley} {et~al.}(2009){Kley}, {Bitsch}, \&
  {Klahr}}]{2009A&A...506..971K}
{Kley}, W., {Bitsch}, B., \& {Klahr}, H. 2009, \aap, 506, 971

\bibitem[{{Kley} \& {Crida}(2008)}]{2008A&A...487L...9K}
{Kley}, W. \& {Crida}, A. 2008, \aap, 487, L9

\bibitem[{{Kley} \& {Lin}(1992)}]{1992ApJ...397..600K}
{Kley}, W. \& {Lin}, D.~N.~C. 1992, \apj, 397, 600

\bibitem[{{Kokubo} \& {Ida}(1998)}]{1998Icar..131..171K}
{Kokubo}, E. \& {Ida}, S. 1998, Icarus, 131, pp.171

\bibitem[{{Kretke} \& {Lin}(2007)}]{2007ApJ...664L..55K}
{Kretke}, K. \& {Lin}, D.~N.~C. 2007, ApJ, 664, L55

\bibitem[{{Lambrechts} \& {Johansen}(2012)}]{2012A&A...544A..32L}
{Lambrechts}, M. \& {Johansen}, A. 2012, A\&A, 544, id.A32

\bibitem[{{Lega} {et~al.}(2014){Lega}, {Crida}, {Bitsch}, \&
  {Morbidelli}}]{Lega2013}
{Lega}, E., {Crida}, A., {Bitsch}, B., \& {Morbidelli}, A. 2014, astro-ph.EP,
  1402.2834

\bibitem[{{Levermore} \& {Pomraning}(1981)}]{1981ApJ...248..321L}
{Levermore}, C.~D. \& {Pomraning}, G.~C. 1981, \apj, 248, 321

\bibitem[{{Levison} {et~al.}(2010){Levison}, {Thommes}, \&
  {Duncan}}]{2010AJ....139.1297L}
{Levison}, H.~F., {Thommes}, E., \& {Duncan}, M.~J. 2010, AJ, 139, pp.1297

\bibitem[{{Lin} \& {Papaloizou}(1986)}]{1986ApJ...307..395L}
{Lin}, D.~N.~C. \& {Papaloizou}, J. 1986, \apj, 307, 395

\bibitem[{{Martin} \& {Lubow}(2014)}]{2014MNRAS.437..682M}
{Martin}, R.~G. \& {Lubow}, S.~H. 2014, MNRAS, 437, p.682

\bibitem[{{Masset} {et~al.}(2006){Masset}, {Morbidelli}, {Crida}, \&
  {Ferreira}}]{2006ApJ...642..478M}
{Masset}, F.~S., {Morbidelli}, A., {Crida}, A., \& {Ferreira}, J. 2006, \apj,
  642, 478

\bibitem[{{Mihalas} \& {Weibel Mihalas}(1984)}]{1984frh..book.....M}
{Mihalas}, D. \& {Weibel Mihalas}, B. 1984, {Foundations of radiation
  hydrodynamics} (New York: Oxford University Press, 1984)

\bibitem[{Morbidelli {et~al.}(2008)Morbidelli, Crida, Masset, \&
  Nelson}]{2008A&A...478..929M}
Morbidelli, A., Crida, A., Masset, F., \& Nelson, R. 2008, A\&A, 478, 929

\bibitem[{{Morbidelli} \& {Nesvorny}(2012)}]{2012A&A...546A..18M}
{Morbidelli}, A. \& {Nesvorny}, D. 2012, A\&A, 546, id.A18

\bibitem[{{Nakagawa} {et~al.}(1986){Nakagawa}, {Sekiya}, \&
  {Hayashi}}]{1986Icar...67..375N}
{Nakagawa}, Y., {Sekiya}, M., \& {Hayashi}, C. 1986, Icarus, 67, p. 375

\bibitem[{{Nelson} {et~al.}(2013){Nelson}, {Gressel}, \&
  {Umurhan}}]{2013MNRAS.435.2610N}
{Nelson}, R.~P., {Gressel}, O., \& {Umurhan}, O.~M. 2013, MNRAS, 435, p.2610

\bibitem[{{Paardekooper} {et~al.}(2010){Paardekooper}, {Baruteau}, {Crida}, \&
  {Kley}}]{2010MNRAS.401.1950P}
{Paardekooper}, S.~J., {Baruteau}, C., {Crida}, A., \& {Kley}, W. 2010, \mnras,
  401, 1950+

\bibitem[{{Paardekooper} {et~al.}(2011){Paardekooper}, {Baruteau}, \&
  {Kley}}]{2011MNRAS.410..293P}
{Paardekooper}, S.~J., {Baruteau}, C., \& {Kley}, W. 2011, MNRAS, 410, 293

\bibitem[{{Paardekooper} \& {Mellema}(2006)}]{2006A&A...459L..17P}
{Paardekooper}, S.~J. \& {Mellema}, G. 2006, \aap, 459, L17

\bibitem[{{Perez-Becker} \& {Chiang}(2011)}]{2011ApJ...727....2P}
{Perez-Becker}, D. \& {Chiang}, E. 2011, ApJ, 727, id.2

\bibitem[{Pierens \& Nelson(2010)}]{2010A&A...520A..14P}
Pierens, A. \& Nelson, R.~P. 2010, A\&A, 520, id.A14

\bibitem[{{Pollack} {et~al.}(1996){Pollack}, {Hubickyj}, {Bodenheimer},
  {Lissauer}, {Podolak}, \& {Greenzweig}}]{1996Icar..124...62P}
{Pollack}, J.~B., {Hubickyj}, O., {Bodenheimer}, P., {et~al.} 1996, Icarus,
  124, 62

\bibitem[{{Shakura} \& {Sunyaev}(1973)}]{1973A&A....24..337S}
{Shakura}, N.~I. \& {Sunyaev}, R.~A. 1973, \aap, 24, 337

\bibitem[{{Takeuchi} \& {Lin}(2002)}]{Takeuchi2002}
{Takeuchi}, T. \& {Lin}, D.~N.~C. 2002, ApJ, 581, pp.1344

\bibitem[{Urpin(1984)}]{1984SvA....28...50U}
Urpin, V. 1984, Soviet Astronomy, 280, p.50

\bibitem[{{Ward}(1997)}]{1997Icar..126..261W}
{Ward}, W.~R. 1997, Icarus, 126, 261

\bibitem[{{Yang} \& {Menou}(2010)}]{2010MNRAS.402.2436Y}
{Yang}, C.-C. \& {Menou}, K. 2010, MNRAS, 402, pp. 2436

\end{thebibliography}
\end{document}